\documentclass[letter]{aa} 
\usepackage{graphicx}
\usepackage{txfonts}
\usepackage[lofdepth,lotdepth]{subfig}
\usepackage{multirow}
\usepackage{chngcntr} 

\counterwithin{paragraph}{subsubsection} 

\begin{document}

   \title{A non-LTE radiative transfer model to study ionized outflows and disks. The case of MWC349A.}


   \author{A. B\'aez-Rubio
          \inst{1}
	  \and J. Mart\'in-Pintado\inst{1}
          \and C. Thum\inst{2}
	  \and P. Planesas\inst{3}
          }

   \institute{Centro de Astrobiolog\'ia (CSIC-INTA),
              Ctra de Torrej\'on a Ajalvir, km 4, 28850 Torrej\'on de Ardoz, Madrid, Spain\\
              \email{baezra@cab.inta-csic.es, jmartin@cab.inta-csic.es}
         \and
		Instituto de Radio Astronom\'ia Milim\'etrica (IRAM), Avenida Divina Pastora, 7, N\'ucleo Central, E 18012 Granada, Spain\\  \email{thum@iram.es}
	\and 
		Observatorio Astron\'omico Nacional (IGN), Alfonso XII 3, E-28014 Madrid, Spain \\ \email{p.planesas@oan.es} 
             }

   \date{Received March 27, 2012; accepted April 28, 2012}

 
  \abstract
   {The best example of a massive star with an ionized outflow launched from its photoevaporating disk is MWC349A. The large amount of reported radio-continuum and radio-recombination line observations toward this galactic UC-HII region offers a unique possibility to build a model of the ionized envelope of this source.} 
{To understand the physical conditions and kinematics of the ionized region of the circumstellar disk and also of the outflow of MWC349A.} 
{We compared the bulk of radio-continuum maps, radio-recombination line profiles, and the H30$\alpha$ centroid map published to date with the predictions of our non-LTE 3D radiative transfer model, MORELI (MOdel for REcombination LInes), which we describe here in detail.}
{Our non-LTE 3D radiative transfer model provides new evidence that the UC-HII region of MWC349A is composed of an ionized circumstellar disk rotating in Keplerian fashion around a star of 38 $\mathrm{M}_{\odot}$, and an ionized outflow expanding with a terminal velocity of 60 km s$^{-1}$ and rotating in the same sense as the disk. The model shows that while maser amplification is the dominant process involved for Hn$\alpha$ radio-recombination line (RRL) emission with quantum numbers $n<41$, stimulated emission is relevant for the emission of RRLs with $n>41$ up at least the H76$\alpha$ line.}
{For the first time, we present a model of MWC349A which satisfactorily explains the vast amount of reported observational data for a very wide range of frequencies and angular resolutions.} 



   \keywords{Stars: massive -- Masers -- 
	HII regions --
	Stars: winds, outflows --
	Accretion, accretion disks
               }

   \titlerunning{A non-LTE model to study ionized envelopes. The case of MWC349A.}
   \authorrunning{B\'aez-Rubio et al.}
   \maketitle
%
\section{Introduction}



Although massive stars (with masses higher than about 8 $\mathrm{M}_{\odot}$) play a key role in the evolution of the Universe, their late stages of evolution as well as the mechanisms that lead to their formation and early evolution are still poorly known (\cite{Zinnecker2007}). Massive stars ionize their surroundings at the early stages of their evolution and also at some evolved stages (such as supergiant B[e] stars) forming Ultra-Compact HII regions (UC-HII). Statistical analysis of the number of UC-HII regions has shown that these objects seem to be confined for longer times than those derived from their crossing times, if it is assumed that they constantly expand at the speed of sound (Churchwell et al. 1990). This raises the question of why high-pressure HII regions do not quickly expand into their surrounding environments of lower pressure.  

Hollenbach et al. (1994) proposed that the long lifetime of UC-HII regions can be explained if they are formed by photoevaporation of the neutral material from the rotating accretion disk left after star formation. This hypothesis would be consistent with two observational results: i) the substantial fraction of UC HII regions ($\sim$30\%) characterized by broad hydrogen-radio recombination lines (with linewidths of $70-200$ km s$^{-1}$), which are interpreted as arising from low velocity ionized outflows associated to neutral disks (Jaffe \& Mart\'in-Pintado, 1999); and ii) the high spatial resolution observations of massive protostars, which have shown that massive star formation seems to proceed, at least in some cases, through accretion of material from a circumstellar disk (\cite{Nielbock2007}, \cite{Jimenez-Serra2009}, \cite{Kraus2010}, \cite{Davies2010}, \cite{Preibisch2011}). Evidence of circumstellar disks have also been found toward evolved massive stars such as supergiant B[e] stars (\cite{{Miroshnichenko2005}}). However, the kinematics of circumstellar disks and ionized envelopes around pre- and post-main sequence massive stars are still poorly understood.





Since its discovery, the UC-HII region of MWC349A has turned out to be a key object for the study of the physical and kinematical properties of these regions thanks to its peculiar characteristics. The star MWC349A is the most intense radio-continuum source at centimeter wavelengths (\cite{Braes1972}) and also one of the most intense sources in the NIR and mid-IR (\cite{Tafoya2004}). The position of this source in the HR diagram clearly corresponds to a massive star (Kraus 2009). According to its spectral features, it is classified as a B[e] star, which are characterized by their forbidden emission lines ([Fe II] and [O I]) in the optical spectrum and its NIR excesses (\cite{Lamers1998}). Nevertheless, its evolutionary stage, as often happens with B[e] stars, is under dispute because of the difficulty caused by the overlap of the pre- and post-main sequence B[e] evolutionary tracks in terms of effective temperature and luminosity (Marston \& McCollum 2008) and the lack of optical absorption lines (\cite{Hamann1988}, \cite{Andrillat1996}). However, it is one of the few massive stars with a well-established circumstellar disk (\cite{White1985} and \cite{Danchi2001}) and an ionized outflow expanding at nearly constant velocity (\cite{Olnon1975}, \cite{Tafoya2004}).

The most peculiar characteristic of MWC349A that allows us to study in detail the kinematics of its ionized regions is that, until the detection toward $\eta$ Carinae (\cite{Cox1995}), Cepheus A HW2 (\cite{Jimenez-Serra2011}) and MonR2-IRS2 (\cite{Jimenez-Serra2013}), it was the only known celestial source with hydrogen radio-recombination line (RRL) masers at millimeter wavelengths (\cite{Martinpintado1989a}). The strong maser amplification of mm and submm RRLs and the strong radio-continuum emission of MWC349A offer a unique possibility to obtain spectra and images with a high spectral and spatial resolution, down to 5 mas (Planesas et al. 1992, \cite{Martinpintado2011}). The comparison of the H30$\alpha$ radio-recombination maser line profile, the H30$\alpha$ centroid map and the free-free radio-continuum emission with the predictions from a non-local thermodynamic equilibrium (non-LTE) 3D radiative transfer MOdel for REcombination LInes (hereafter MORELI), has provided stringent constraints on the disk and outflow kinematics of MWC349A as well as its electron temperature and electron density distribution (\cite{Martinpintado2011}). The observations were explained by assuming two kinematical components, a thin virtually edge-on Keplerian ionized disk around a massive star ($\sim$ 30-60 $\mathrm{M}_{\odot}$), responsible for the maser spikes observed in the H30$\alpha$, and a rotating and expanding ionized wind with a terminal velocity of about 70 km s$^{-1}$. It also led to the remarkable conclusion that the outflow seems to be launched from the Keplerian disk at a radius of $\le $20 AU from the star. Magnetic wind models (both X-wind models, \cite{Shu1994}, and disk wind models, \cite{Blandford1982}) seem to be the most suitable to explain the region where the wind emerges and accelerates up to its terminal velocity. These results are extremely important since they provide clues on the origin of ionized winds from massive stars with circumstellar disks at very small spatial scales.



This paper gathers all the centimeter, millimeter, submillimeter, and far-infrared observations of the continuum and RRLs toward MWC349A to establish, for the first time, a model that reproduces the bulk of the observed features found over this wide frequency range. We carried out a thorough analysis of all the data by using a 3D non-LTE radiative transfer model to constrain the structure, physical conditions, and the kinematics. In Sect. \ref{section_model}, we discuss the basic physics incorporated in the MORELI radiative transfer model required for the interpretation of both the continuum and RRL emission. In particular, we describe the line formation under non-LTE conditions and the effects of saturation of masers. In Sect. \ref{physical_cond_and_kin}, we describe the different geometries, physical and kinematical structure included in MORELI. Then, in Sect. 4, we report the first model for MWC349A that consistently explain most of the observational data available to date. Also, in Sect. 4, we show how the main parameters of the ionized gas were constrained by comparing the observations with the predictions of MORELI. While in Sect. 4 we focus on the H30$\alpha$ centroid map and its line profile, as well as the integrated-line intensities of RRLs from cm to the mid-IR wavelengths, in Sect. \ref{section_profiles} we show the fit of our predictions for the rest of RRLs observed so far. Finally, conclusions are drawn in Sect. \ref{section_conclusions}.




\section{Non-LTE radiative transfer model} 
\label{section_model}

\subsection{Radiative transfer integration}



The resulting spectrum for every line-of-sight is obtained by discretizing the sources in regular 3-D grid cells with sizes of $dx$, $dy$, and $dz$, where the $z$ axis refers to the axis along the line-of-sight (with the observer toward negative $z$ values), $x$ the revolution symmetry axis projected into the plane of the sky, and $y$ the axis orthogonal to the $x\mathrm{-}z$ plane. Thus, this discretization observed in the plane of the sky, $\left(x,y\right)$, is equivalent to a mesh of points. We solve the radiative transfer equation for every cell along the line-of-sight by assuming constant properties in the whole cell volume

\begin{equation}
\frac{dI_{\nu}}{d\tau_{\nu}}=-I_{\nu}+S_{\nu} \mathrm{,}
\label{equation:transfer_rad_eq1}
\end{equation}

\noindent where $I_{\nu}$ is the specific radiation intensity at the frequency $\nu$, $S_{\nu}$ is the source function, and $\tau_{\nu}$ is the optical depth (defined as $d\tau_{\nu}=\kappa_{\nu} dl$, with $\kappa_{\nu}$ the absorption coefficient of the cell with physical length $l$).

We consider, as an approximation, that the electron density of the source is large enough to assume that the diffuse radiation is locally absorbed leading to a subsequent photoionization, the so-called on-the-spot approximation (\cite{Osterbrock1989}). The outgoing radiation of cell $i$ for a given frequency, $\nu$, is then given by

\begin{equation}
I_{\nu, i}=I_{\nu, {i-1}} \mathrm{e}^{-\tau_{\nu, i}}+B_{\nu}(T_{\mathrm{e}})(1-\mathrm{e}^{-\tau_{\nu, i}}) \mathrm{,}
\label{equation:transfer_rad_eq}
\end{equation}

\noindent where $I_{\nu,{i-1}}$ is the background emission (which is given by the cosmic microwave background radiation at the rear side of the source where $i=1$), $B_{\nu}(T_{\mathrm{e}})$ the source function (given by Planck's law for a black body with electron temperature $T_\mathrm{e}$), and $\tau_{\nu,i}$ the optical depth from the cell $i$. 

\subsubsection{Continuum emission}

The optical depth for the free-free continuum radiation is calculated by using the absorption coefficient

\begin{eqnarray}
\label{continuum_optical_depth}
\kappa_{\nu \mathrm{, c}}&=&\frac{4\sqrt{2\pi}}{3\sqrt{3}}\frac{1}{c \left(m_\mathrm{e} k T_\mathrm{e}\right)^{3/2}}\left(\frac{e^2}{4\pi \epsilon_0}\right)^3 \frac{N_\mathrm{e} N_\mathrm{i}}{\nu^2}\left(g_{\mathrm{ff},\nu}+g_{\mathrm{bf},\nu}\right) \\
&\approx& 1.77\times 10^{-12} \frac{N_\mathrm{e} N_\mathrm{i} \left(g_{\mathrm{ff},\nu}+g_{\mathrm{bf},\nu}\right)}{\nu^2 T_\mathrm{e}^{3/2}}\ \ \ \mathrm{[SI\ units]} \quad \textrm{if  \  } h \nu \ll k T_\mathrm{e} \mathrm{,} \nonumber 
\end{eqnarray}



\noindent where $m_\mathrm{e}$ and $e$ are the electron mass and charge, $h$ and $k$ the Boltzmann and Planck constants, $c$ the speed of light, $\epsilon_0$ the vacuum permittivity, $N_\mathrm{e}$ and $N_\mathrm{i}$ the electron and ionic densities and $g_{\mathrm{ff},\nu}$ and $g_{\mathrm{bf},\nu}$ the Gaunt factors for free-free and bound-free transitions, respectively (see Appendix \ref{gaunt_appendix}).

\subsubsection{LTE RRL emission}

In the formation of RRLs, one has to simultaneously consider the contribution of both the continuum and line emission. According to the quantum theory of the radiation, spontaneous emission, stimulated emission, and absorption are the three basic processes through which photons and atoms interact (\cite{Einstein1916}). The probability per unit time for an electron to spontaneously decay from the upper level $m$ to the lower level $n$ is given by the Einstein coefficient of spontaneous emission, $A_{mn}$ (\cite{Towle1996}). On the  other hand, we use the spectral energy density, $u_\nu$, to define the Einstein B-coefficients. The transition probability per unit time for absorption and stimulated emission are $P_{\mathrm{a}}=u_\nu B_{nm}$ and $P_{\mathrm{st}}=u_\nu B_ {mn}$, respectively, where $u_\nu$ is related to $I_\nu$

\begin{equation}
u_\nu=\frac{1}{c} \int I_{\nu} \, d\Omega \mathrm{,}
\end{equation}

\noindent and where $\Omega$ is the solid angle. 

Using the Einstein coefficients described previously and treating the stimulated emission as a negative absorption, the line absorption coefficient depends on the electron population of the involved levels, $N_m$ and $N_n$, according to the expression


\begin{equation}
\kappa_{\nu \mathrm{,l}}=\frac{h \nu}{c} \left(1-\frac{g_n}{g_m}\frac{N_m}{N_n} \right) N_{n} B_{nm} \Phi_{\nu} \mathrm{,}
\label{Einstein}
\end{equation}

\noindent where $B_{nm}$ is the Einstein coefficient for absorption, $g_n$ and $g_m$ are the statistical weights of the electronic levels $n$ and $m$ respectively (with $g_k=2 k^2$ for $k=n,m$), and $\Phi_{\nu}$ the line profile (see Appendix \ref{RRL_profiles}).


The values of $N_m$ and $N_n$ are determined by solving the equation system of statistical equilibrium considering all the processes involved in the excitation of electronic levels. Under LTE conditions, collisions dominate over the radiative processes and the electronic levels are populated according to the Saha-Boltzmann distribution. In this case, Eq. \ref{Einstein} yields:






\begin{eqnarray}
\kappa_{\nu \mathrm{,l}} & = &\frac{h \nu}{c} \left(1-\mathrm{e}^{-\frac{h \nu}{k T_e}}\right)N_\mathrm{e} N_\mathrm{i} \frac{g_n}{2} \left(\frac{h^2}{2\pi m_\mathrm{e} k T_\mathrm{e}}\right)^{3/2}\exp{\left(\frac{h \nu_0}{k T_\mathrm{e}}\right)} B_{nm} \Phi_{\nu} \nonumber \\
& \approx & 7\times 10^{-70} \left(1-\mathrm{e}^{-\frac{h \nu}{k T_e}}\right)N_\mathrm{e} N_\mathrm{i}  n^2  \mathrm{e}^{\frac{h \nu_0}{k T_\mathrm{e}}} \nu B_{nm} \Phi_{\nu} \mathrm{\ \ [SI\ units]} \mathrm{,}
\end{eqnarray}

\noindent where $B_{nm}$ is estimated according to the Dirac theory described by Towle et al. (1999).





\subsubsection{Non-LTE RRL emission}
\label{departure_coefficients}

The absorption coefficient discussed in the previous section must be modified for the case of non-LTE since the electron population departs from that given by the Saha-Boltzmann distribution as a result of radiative transitions dominating over collisional transitions. To account for this one introduces for each electronic level, $n$, a correction factor known as the departure coefficient, $b_n$, that relates the electron population in the non-LTE case, $N_n$, with the LTE case, $N_n^*$,

\begin{equation}
N_n=b_n N_n^* \mathrm{.}
\end{equation}



\begin{figure}
\centering
\includegraphics[angle=0,width=8cm]{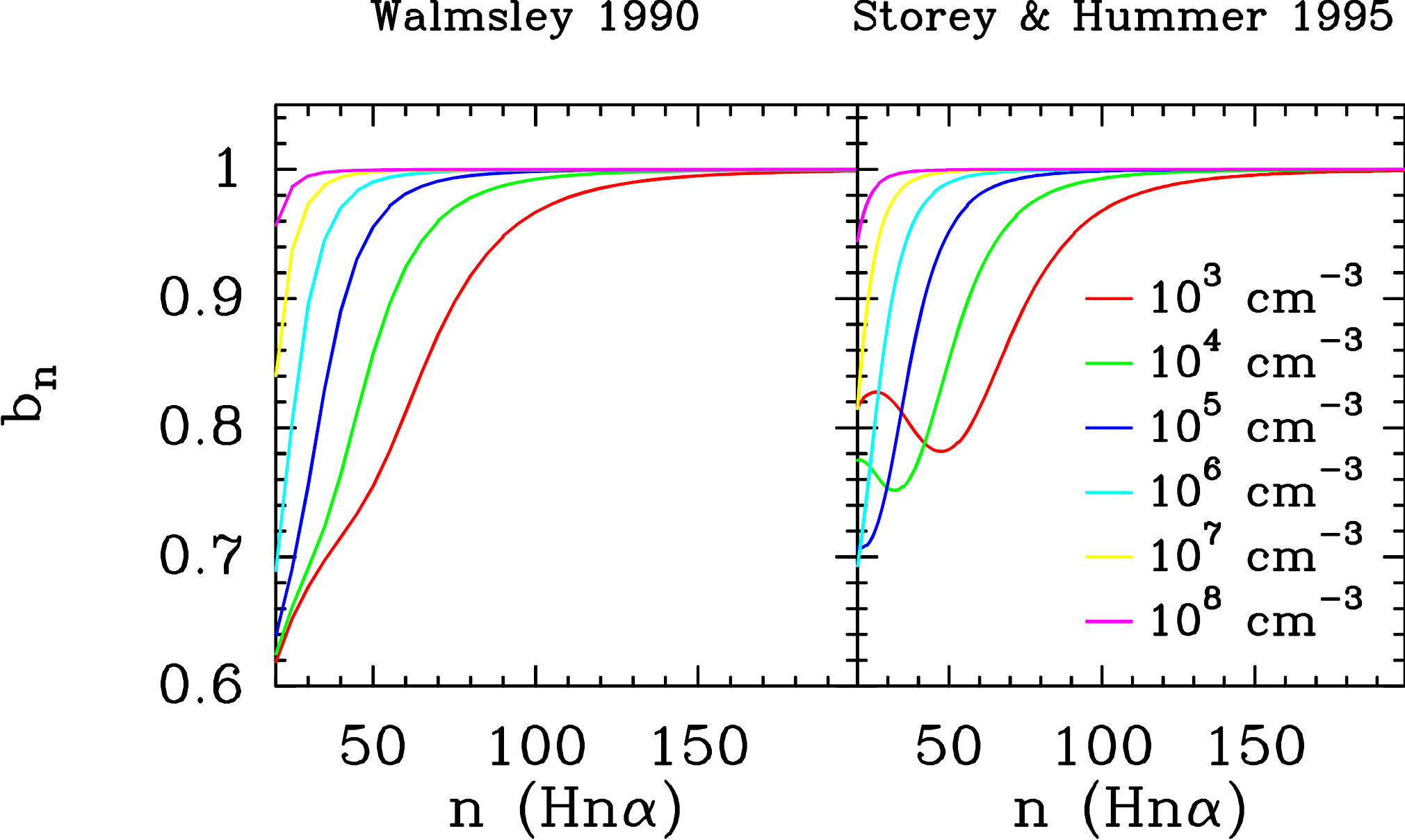}
\caption{Estimated departure coefficients, $b_n$, for Hn$\alpha$ lines, an electron temperature of 10000K, and different electron densities by \cite{Walmsley1990} and \cite{Storey1995}.}
\label{bn_coefficients}
\end{figure}

These departure coefficients depend on both collisional and radiative processes. They have been computed independently by several authors taking into consideration different ranges of electron density and electron temperature. To date, the two most extensive $b_n$ coefficient tables are those reported by \cite{Storey1995} and \cite{Walmsley1990}. In Fig. \ref{bn_coefficients} we show their values for an ionized gas of 10000 K for different electron density ranges. The MORELI code incorporates the parametrization of the $b_n$ values obtained in both studies for the case of an optically thin HII-region for all RRLs except for the Lyman lines. We remark that this procedure will give us approximate results since the departure coefficients would actually depend on the radiation field for our particular assumed geometry and physical structure, which will differ from those assumed in the models used in \cite{Storey1995} and \cite{Walmsley1990}.

As described by Dupree \& Goldberg (1970), the RRL absorption coefficient can be then expressed as $\kappa_{\nu \mathrm{,l}}=\kappa_{\nu \mathrm{,l}}^* b_n \beta_{mn}\ $ in the non-LTE case, where $\beta_{mn}$ is defined



\begin{equation}
\beta_{mn}=\frac{1-\frac{b_m}{b_n}e^{-\frac{h\nu}{k T_\mathrm{e}}}}{1-e^{-\frac{h \nu}{k T_\mathrm{e}}}} \mathrm{.}
\label{beta_equation}
\end{equation}



\begin{figure}
\centering
\includegraphics[angle=0,width=8cm]{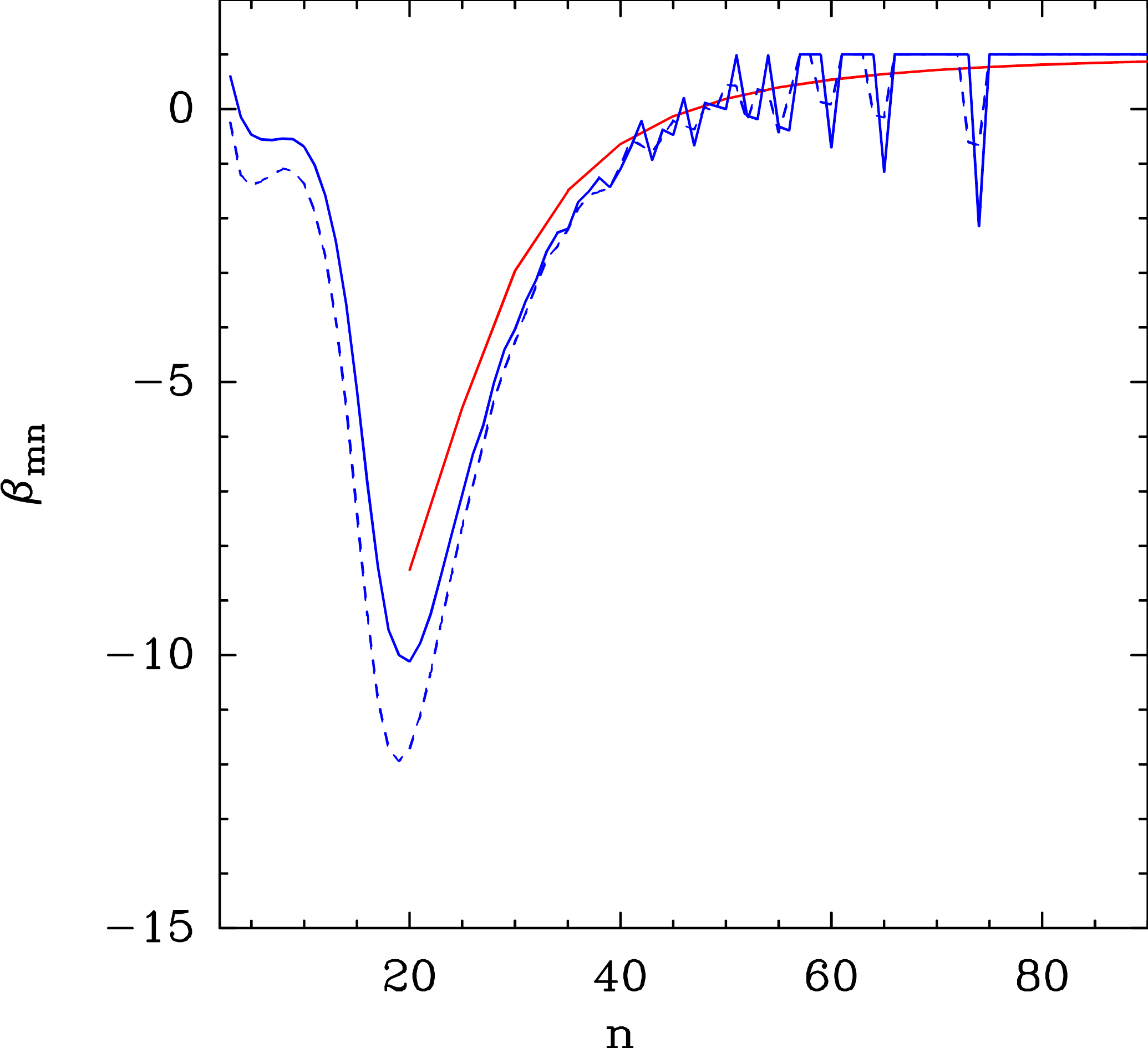}
\caption{Estimated $\beta_{mn}$ coefficients, $b_n$ for an ionized gas with electron density of $10^7$ cm$^{-3}$, and electron temperature of 10000 K. The solid red line shows the $\beta_{mn}$ values derived by \cite{Walmsley1990} for Hn$\alpha$ transitions, while the solid and dashed blue lines show the values derived by \cite{Storey1995} for Hn$\alpha$ and Hn$\beta$ transitions, respectively. The observed zig-zag for the $\beta_{mn}$ \cite{Storey1995} coefficients for $n>$ is due to the numerical error in deriving them from Eq. \ref{beta_equation}. This error increases for increasing $n$ since the relative difference between the $b_{m}$ and $b_{n}$ decreases as observed in Fig. 1.}
\label{beta_coefficients}
\end{figure}

For certain combinations of $n$, $T_\mathrm{e}$, and $N_\mathrm{e}$ found in some UC-HII regions (i.e. in MWC349A and Cep A HW2), the level populations can be inverted and, subsequently, the second term of the latter equation can be larger than unity. It implies that $\beta_{mn}< 0$ and, therefore, the incident radiation is amplified. We can see in Fig. \ref{beta_coefficients} that this is the case for an ionized gas with electron density of $10^7$ cm$^{-3}$ and electron temperature of 10000 K as found in MWC349A.

Using the above mentioned departure coefficients, it is possible to derive the non-LTE source function. And from the latter, by integrating the radiative transfer equation, one obtains the brightness temperature at the center of the line (\cite{Dupree1970})

\begin{equation}
T_{B}=T_\mathrm{bg} \mathrm{e}^{-\left(\tau_{\nu\mathrm{,l}}+\tau_{\mathrm{\nu,c}}\right)}+T_\mathrm{e}\left(\frac{\tau_{\nu\mathrm{,l}}}{\beta_{mn}}+\tau_{\nu\mathrm{,c}}\right) \frac{1}{\tau_{\nu\mathrm{,l}}+\tau_{\nu\mathrm{,c}}}\left(1-\mathrm{e}^{-\left(\tau_{\nu \mathrm{,l}}+\tau_{\nu\mathrm{,c}}\right)}\right) \mathrm{,}
\label{equation_line_non_LTE}
\end{equation}

\noindent where $T_\mathrm{bg}$ is the background temperature respectively. This equation reduces to Eq. \ref{equation:transfer_rad_eq} for the LTE case ($b_n=1$).



\subsection{Stimulated and maser emission}

Soon after the discovery of RRLs, it was noticed that for the typical physical conditions found in ionized regions, the RRLs should be emitted under non-LTE conditions (\cite{Goldberg1966}), mainly stimulated emission. This type of emission is due to the amplification of the background continuum and line emission when the optical depth is negative but larger than $-1$, that is, $-1<\tau_{\nu \mathrm{,c}}+\tau_{\nu \mathrm{,l}}<0$. Nevertheless, if the pumping mechanisms producing the inversion of population in the ionized gas are intense enough, the amplification of the radiation can become very intense, $\tau_{\nu \mathrm{,c}}+\tau_{\nu \mathrm{,l}} \ll -1$. This is referred to as maser amplification. In this case, the line intensity (Eq. \ref{equation_line_non_LTE}) would have an approximately exponential dependence on optical depth and, therefore, on physical conditions (electron density and electron temperature)

\begin{equation}
|T_\mathrm{B}(\tau_{\nu \mathrm{,c}}+\tau_{\nu \mathrm{,l}})|\propto \mathrm{e}^{|\tau_{\nu \mathrm{,c}}+\tau_{\nu \mathrm{,l}}|} \mathrm{.}
\end{equation}


Nevertheless, the exponential growth of the intensity along a line of sight with the required physical conditions for maser amplification cannot continue indefinitely. This is because the efficiency for stimulated emission reaches an upper limit when each pump event results in the release of a maser photon (\cite{Strelnitski1996b}). In this case the intensity increases linearly with the optical depth and it is referred to as a saturated maser.




\subsection{Saturation effects}
\label{saturation_section}

The full treatment of saturation effects should take into account the changes in the electron populations (the $b_n$ coefficients) induced by the maser radiation from all directions. This full treatment is clearly out of the scope of this paper. As a first approximation, we can estimate the saturation effects by considering only the changes in $b_n$ coefficients caused by the incident radiation in the line-of-sight. From the resolution of the equations of statistical equilibrium under a particular set of approximations, we obtain the absorption coefficient for the saturated case (\cite{Strelnitski1996b})





\begin{equation}
\kappa_{\nu \mathrm{,l, sat}}=\frac{\kappa_{\nu\mathrm{,l}}}{1+J_{\nu}/J_{\nu \mathrm{, sat}}} \mathrm{,}
\label{saturation_equation}
\end{equation}

\noindent where $\kappa_{\nu}$ is the absorption coefficient for the unsaturated regime ($J_{\nu} \ll J_{\nu \mathrm{, sat}}$), $J_{\nu}$ is the intensity averaged over the whole solid angle ($J_{\nu}=\frac{1}{4\pi} \int I_{\nu} \, d\Omega $), and $J_{\mathrm{sat}}$ is the saturation intensity, which is defined as

\begin{equation}
J_{\nu \mathrm{, sat}}=\frac{2 h \nu_{0}^3}{c^2} \frac{C_{\mathrm{t}}+C_{mn}}{A_{mn}} \mathrm{,}
\label{saturacion}
\end{equation}

\noindent with $C_{\mathrm{t}}$ the collisional coefficient from the maser levels to other levels and $C_{mn}$ the collisional coefficient between the two involved maser levels. The sum $C_t+C_{mn}$ is derived from Table \ref{table:pressure_broadening} ($C_{\mathrm{t}}+C_{mn}=\Delta\nu_\mathrm{l}/ \left(2\pi\right)$).

For strong maser emission ($J_{\nu} \gg J_{\nu \mathrm{, sat}}$), the absorption coefficient given in Eq. \ref{saturation_equation} becomes $\kappa_{\nu \mathrm{, sat}} \approx \kappa_{\nu} J_{\nu \mathrm{, sat}}/J_{\nu}$. Substituting this into the radiative transfer Eq. \ref{equation:transfer_rad_eq1} results in an equation whose integration gives a linear dependence of the outgoing intensity with the optical depth

\begin{equation}
T_{\mathrm{B}}= \left(T_{\mathrm{e}}\left(\frac{\tau_{\nu\mathrm{,l}\mathrm{,\ sat}}}{\beta_{mn}}+\tau_{\nu\mathrm{,c}}\right) \frac{1}{\tau_{\nu\mathrm{,l}\mathrm{, sat}}+\tau_{\nu\mathrm{,c}}}-T_{\mathrm{sat}}\right) (\tau_{\nu\mathrm{,l} \mathrm{,\ sat}}+\tau_{\nu\mathrm{,c}}) \mathrm{,}
\label{intensity_saturation_case}
\end{equation}

\noindent where $\tau_{\nu \mathrm{,l} \mathrm{,\ sat}}$ is the line optical depth taking into account the saturation effects (see Eq. \ref{saturation_equation}) and $T_{\mathrm{sat}}$ the saturation temperature. 


To estimate the saturation effects, we uses MORELI to compute the saturation degree, $J_{\nu \mathrm{, sat}}/J_{\nu}$, assuming a solid angle for the maser beam, $\Omega_{\mathrm{m}}$.

\section{Physical structure and kinematic considerations for the source model}
\label{physical_cond_and_kin}

\subsection{Geometry}
\label{geometria}

The MORELI code offers the possibility to consider different source geometries for the wind, such as a sphere, a cylinder (specified by its radius and length), a double-cone (specified by the semi-opening angle, $\theta_\mathrm{w}$, and length) or a hyperboloid (specified by its minimum width, semi-opening angle and length). In all cases MORELI incorporates the possibility that the strong radiative pressure coming from the central star has swept up all the ionized gas in the region within a radius $r_\mathrm{min}$. For all the different geometries except for the sphere, one also needs to specify the inclination angle, $\theta_i$, of the revolution axis with respect to the plane of the sky.

Although physically an unbounded ionized wind could be extended virtually to infinity, with no well definite boundaries, our model limits the size to a maximum radius, $r_\mathrm{max}$, for all the geometries. This radius is one of the critical parameters in our model to adequately sample most of the flux in the RRLs tracing the outer regions of the ionized outflow. The MORELI code allows this radius to be introduced by the user or to be estimated from the effective radius, $R_{\mathrm{eff}}$, defined to contain a specific percentage of the free-free radio-emission produced for an isothermal spherical ionized wind isotropically expanding at constant velocity with electron density distribution given by $N_{\mathrm{e}}\left(r\right)=N_{\mathrm{e}}\left(r=1\right) r^{-2}$ (see Sect. 3.4). Since we do not consider only this particular wind structure for the UC-HII regions, we essentially trace all the emission by considering that the source has a size $n_\mathrm{R}$ times that of the effective radius, that is


\begin{equation}
r_\mathrm{max}=n_\mathrm{R}\cdot R_{\mathrm{eff}} \mathrm{.}
\end{equation}


For a partially optically thick wind, the derived effective radius containing half of the free-free radio-continuum flux is given (\cite{Panagia1975})

\begin{equation}
R_{\mathrm{eff},\ \nu}\approx 4.79\times10^{20} \ \nu^{-0.7}T_\mathrm{e}^{-0.45}\dot{M}^{2/3}\mu^{-2/3}v_{0}^{-2/3}\ \ \ \mathrm{[SI\ units]} \mathrm{,}
\end{equation}



\noindent where $\mu$ is the main atomic weight per electron and $\dot{M}$ the mass loss rate in units of $\mathrm{M}_{\odot}/\mathrm{year}$. Instead, for an optically thick wind, the effective radius containing the percentage $p$ of the free-free radio-continuum flux results (see Appendix \ref{appendix_effective_radius})

\begin{equation}
\label{radio_efectivo}
R_{\mathrm{eff}}=\frac{\pi}{4\left(1-p\right)}r_{\mathrm{min}} \mathrm{.}
\end{equation}

While the cylinder geometry is adequate to model sources with very collimated radio-jets (i.e. CRL618; \cite{Martinpintado1988}), the double-cone and the hyperboloid geometries are adequate to model sources with wide ionized bipolar outflows.

\subsection{Modelling of the kinematics}
\label{kinematics_modelling}

\subsubsection{Disk kinematics}

\begin{figure}
\centering
\includegraphics[angle=0,width=8cm]{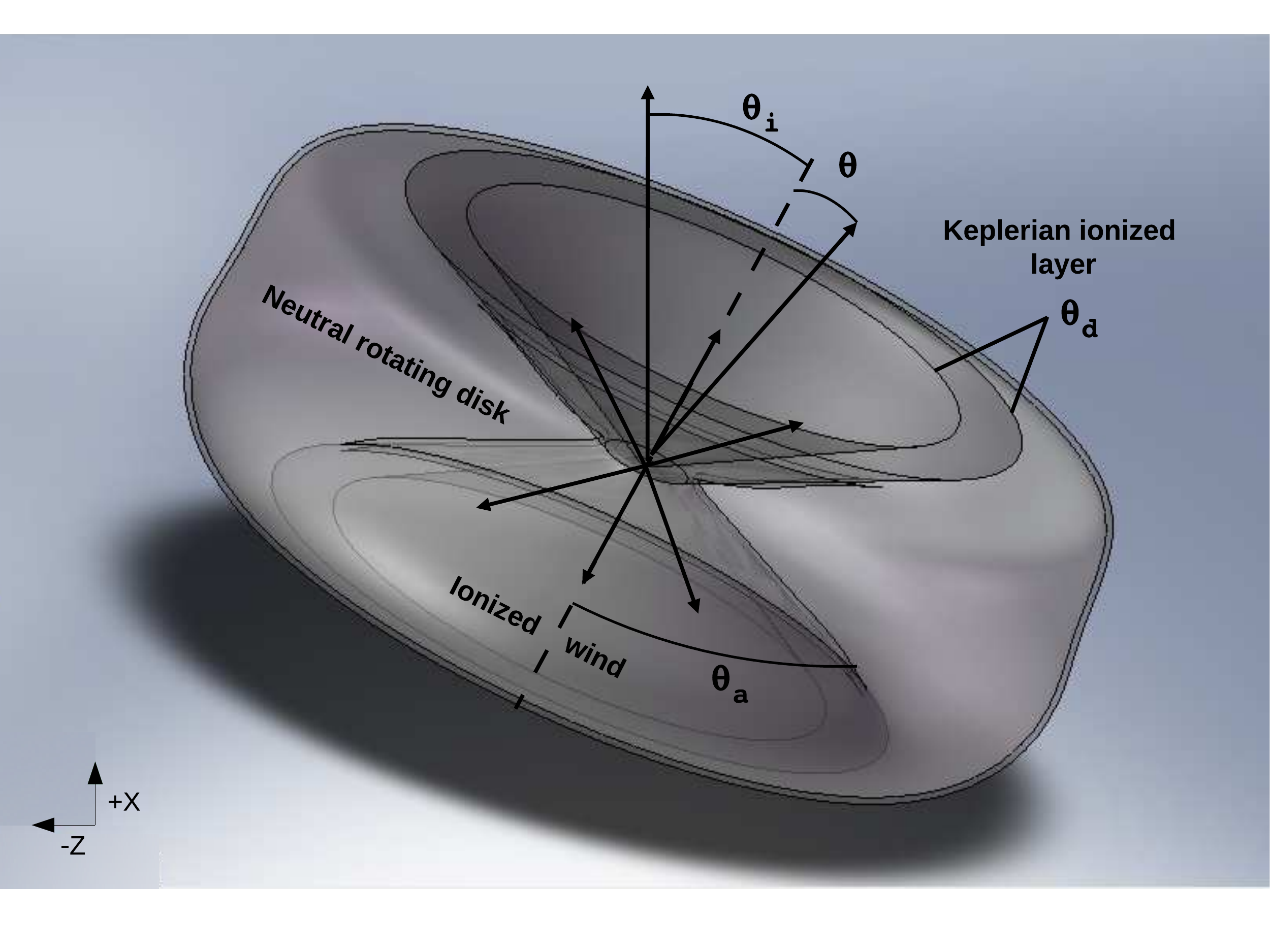}
\caption{Sketch of the double-cone geometry used for the modelling of MWC349A. The ionized gas is contained in a double cone with its revolution axis perpendicular to the plane of the neutral disk. Each cone has a semi-opening angle of $\theta_{\mathrm{a}}$ and is made up of two ionized regions with different kinematics: i) a rotating and expanding ionized wind contained within the double-cone with semi-opening angle $\theta_{\mathrm{a}}-\theta_{\mathrm{d}}$; and ii) a Keplerian rotating ionized layer comprised between the ionized wind and the neutral disk with an opening angle relative to the conical surface of the double cone of $\theta_{\mathrm{d}}$. The observer is located toward the negative $z$-axis. The derived inclination angle $\theta_\mathrm{i}$ is such that the source is tipped up in front.}
\label{figure:figura}
\end{figure}

For the particularly interesting case of a source with a neutral disk and an expanding ionized outflow represented by a double-cone geometry, the model allows us to consider a kinematic component consisting in an ionized Keplerian rotating layer located next to the neutral disk. The ionized layer will be specified by its opening angle relative to the conical surface of the double-cone, $\theta_\mathrm{d}$ (see Fig. \ref{figure:figura}), and the radius up to which the Keplerian rotation extends, $r_\mathrm{d}$. Hereafter, we will refer to this layer as  the ionized Keplerian rotating disk. The velocity rotation component is added to the expansion, being its projection along the line-of-sight {\bf(see Appendix \ref{disk_kinematics}) given by

\begin{eqnarray}
v_z=V_\mathrm{Kepler}\frac{y\cdot \cos(\theta_\mathrm{i})}{\left (y^2+z_{\mathrm{d}}^2 \right )^{3/4}}\\
z_{\mathrm{d}}\equiv \cos(\theta_i) z-\sin(\theta_i)x \nonumber \mathrm{,}
\label{vz_equation}
\end{eqnarray}

\noindent where} $V_\mathrm{Kepler}$ is the Keplerian velocity at a radius equal to unity ($V_\mathrm{Kepler}=\sqrt{G M}$), $M$ the central mass of the star, and $G$ the universal gravitational constant.

\subsubsection{Outflow kinematics}
\label{outflow_kinematics_subsection}

The model also considers different possibilities for the kinematics of the ionized gas depending on the geometry and on the features of the source to model. For all geometries it is possible to consider the kinematics given by the Hollenbach model (\cite{Hollenbach1994}). It is included in a simplified way considering two discontinuous kinematic components, a Keplerian rotating disk confined in its gravitationally bounded ionized atmosphere, and a rotating and expanding wind at the terminal velocity, v$_0$, outside. The radius at which the discontinuity occurs (known as gravitational radius, $R_\mathrm{g}$) is defined as the distance at which the most likely thermal velocity equals the orbital velocity of the Keplerian disk due to the star's gravitational potential. Subsequently, this radius is given by the expression (\cite{Hollenbach1994})

\begin{equation}
R_\mathrm{g}=\frac{G M m_\mathrm{e}}{k T_\mathrm{e}} \mathrm{.}
\end{equation}

The required value for $R_\mathrm{g}$ to explain the observations can be also considered as a free input parameter in the model. The procedure applied is an approximation since there is not a specific radius from which the ionized gas can escape, but a continuous range of radii where the different individual ions with different velocities reach the escape velocity.

Another possibility incorporated in MORELI is to consider a constant acceleration for the wind, reaching its terminal velocity at a radius $r_\mathrm{a}$. This is the case of the modelling of Cep A HW2 (\cite{Jimenez-Serra2011}).

The projected velocity along the line-of-sight for those radii where the wind has reached its terminal velocity ($r>r_\mathrm{a}$) or for the case when it is assumed that the wind reaches the terminal velocity suddenly at $r_\mathrm{a}=0$ is given by

\begin{equation}
\mathrm{v}_z=\mathrm{v}_0\frac{z}{r} \mathrm{.}
\label{vz_equation_with_accel}
\end{equation} 

While for those radii where the wind is still being accelerated ($r<r_\mathrm{a}$) the expression results (see Appendix \ref{outflow_accel})

\begin{equation}
\mathrm{v}_z=\mathrm{v}_0\frac{z}{r_\mathrm{a}} \mathrm{.}
\end{equation} 


Finally, we note that in addition to the expansion, we can also consider another kinematic component for the ionized outflow: its rotation following a Keplerian law similar to that of the disk.

\subsection{Electron temperature and electron density distribution}

The MORELI code provides the possibility of considering electron temperature, $T_\mathrm{o}$, and electron density, $N_\mathrm{e}$, gradients specified by power-law with indices, $b_\mathrm{t}$ and $b_\mathrm{d}$ respectively. The values of $T_\mathrm{o}$ and $N_\mathrm{e}$ are specified in the model at a particular radius. Moreover, for the case of the double-cone geometry, we also incorporate a possible dependence of the electron density on the angle with respect to the revolution axis of the cone, $\theta$ (see Fig. \ref{figure:figura}), since physically we expect the higher density close to the neutral disk if it is photoevaporating. Thus, the electron density distribution assumed in the model is given by

\begin{equation}
\label{Ne_profile}
N_{\mathrm{e}}\left(r,\theta\right)=N_\mathrm{e}\left(r=1,\theta=\theta_\mathrm{a}\right)\frac{\mathrm{e}^{-\left(\theta_\mathrm{a}-\theta\right)/\theta_0}}{r^{b_\mathrm{d}}} \mathrm{,}
\end{equation}

\noindent where $\theta_{\mathrm{a}}\equiv \theta_{\mathrm{w}}+\theta_{\mathrm{d}}$ is the semi-opening angle of the whole ionized gas, $\theta_\mathrm{a}-\theta$ is the angle between an ionized gas cell and the conical surface of the neutral disk (see Fig. \ref{figure:figura}) and $\theta_0$ is a free dimensionless input parameter {used to model the angular electron density dependence} ($\theta_0 = \infty$ if one assumes there is no dependence of $N_\mathrm{e}$ on the angle).

Although Eq. \ref{Ne_profile} refers to both the ionized outflow and ionized circumstellar disk located next to the neutral disk, the model also considers the possibility that this disk has a different electron density, $N_\mathrm{e,d}$, and/or electron temperature, $T_\mathrm{d}$, than the ionized wind to account for the possible difference in cooling due to its higher electron density.

The geometry of the source, and the electron density and electron temperature distributions are constrained by fitting the radio-continuum images and the observed spectral energy distribution (SED). 

\section{The case of MWC349A}
\label{section_MWC349A}

In the following sections we describe in detail how the different observations have been used to constrain the values of the input parameters of the model for the ionized gas in MWC349A. Table \ref{table:inputs} shows the parameters that fit best the overall data set available to date at the radio and millimeter wavelengths. The results presented in this paper have been derived by assuming a double-cone geometry with two kinematical components (ionized outflow and ionized circumstellar disk) and with sudden acceleration of the ionized wind. Since the outflow is partially optically thick, we have assumed the effective radius derived by \cite{Panagia1975} (see Sect. \ref{geometria}). Although the distance for MWC349A is under debate (\cite{Meyer2002}), we have considered the typically assumed distance of 1.2 kpc (\cite{Cohen1985}).

The best fit was obtained by using the $b_n$ coefficients reported by Storey \& Hummer 1995 (see Sect. \ref{section_H30alpha}).

\begin{table*}
\caption{Final input parameters for the best fit to the continuum and RRL data of MWC349A.}
\label{table:inputs}
\begin{tabular}{lccc} \hline 
\textbf{Input parameter}&\textbf{Value}&\textbf{Constrained from}&\\ \hline \hline
Central mass, M&38$\mathrm{M}_\mathrm{sun}$&RRL profiles& star\\ 
Density distribution, $N_\mathrm{e}(r,\theta)$&1.6 $\times 10^9 r^{-2.14} \exp{\left[\left(\theta_\mathrm{a}-\theta\right)/20 \right]}\ \mathrm{cm}^{-3}\ \ \ ^\mathrm{a}$&radio-continuum&\multirow{3}{*}{ \Bigg{\}} disk \& wind }\\ 
Inner radius, $r_\mathrm{min}$&0.05 AU $^\mathrm{b}$&radio-continuum& \\ 
Double-cone's semi-opening, $\theta_\mathrm{a}$ & 57$\degr$&radio-continuum& \\  \hline

Inclination angle, $\theta_\mathrm{i}$&8$\degr$ (tipped up)&H30$\alpha$ centroid map&\multirow{4}{*}{disk}\\ 
Opening angle of the ionized disk, $\theta_\mathrm{d}$&6.5\degr&RRL profiles&\\ 
Ionized disk's outer radius, $r_\mathrm{d}$&130 AU&radio-continuum&\\ 
Ionized disk electron temperature, $T_\mathrm{d}$& 9450 K&RRL profiles&\\  \hline 

Ionized outflow's outer radius, $n_\mathrm{R}$ &1.5$R_{\mathrm{eff,}\nu}$&RRL profiles&\multirow{4}{*}{wind} \\
Outflow electron temperature, $T_\mathrm{o}$&12000 K&radio-continuum&\\ 
Outflow terminal velocity, $\mathrm{v}_{0}$&60 $\mathrm{km\ s}^{-1}$&H30$\alpha$ centroid map and RRL profiles&\\ 
Outflow turbulent velocity, $\mathrm{v}_\mathrm{tu}$&15 $\mathrm{km\ s}^{-1}$&H30$\alpha$ centroid map and RRL profiles&\\ \hline


\end{tabular}

\begin{list}{}{}
\item[$^{\mathrm{a}}$] Where $r$ is in units of 10 AU and $(\theta_\mathrm{a}-\theta)$ in radians.
\item[$^\mathrm{b}$] We assume $r_\mathrm{min}=0.05$ AU even when we can only constrain that the stellar wind reaches radii as inner as 3 AU (Sect. \ref{section_MWC349A}). 
\end{list}

\end{table*}

\subsection{Constraining the physical structure and geometry of MWC349A on the basis of the SED and radio-continuum images} 

\begin{figure}
\centering
\includegraphics[width=9cm]{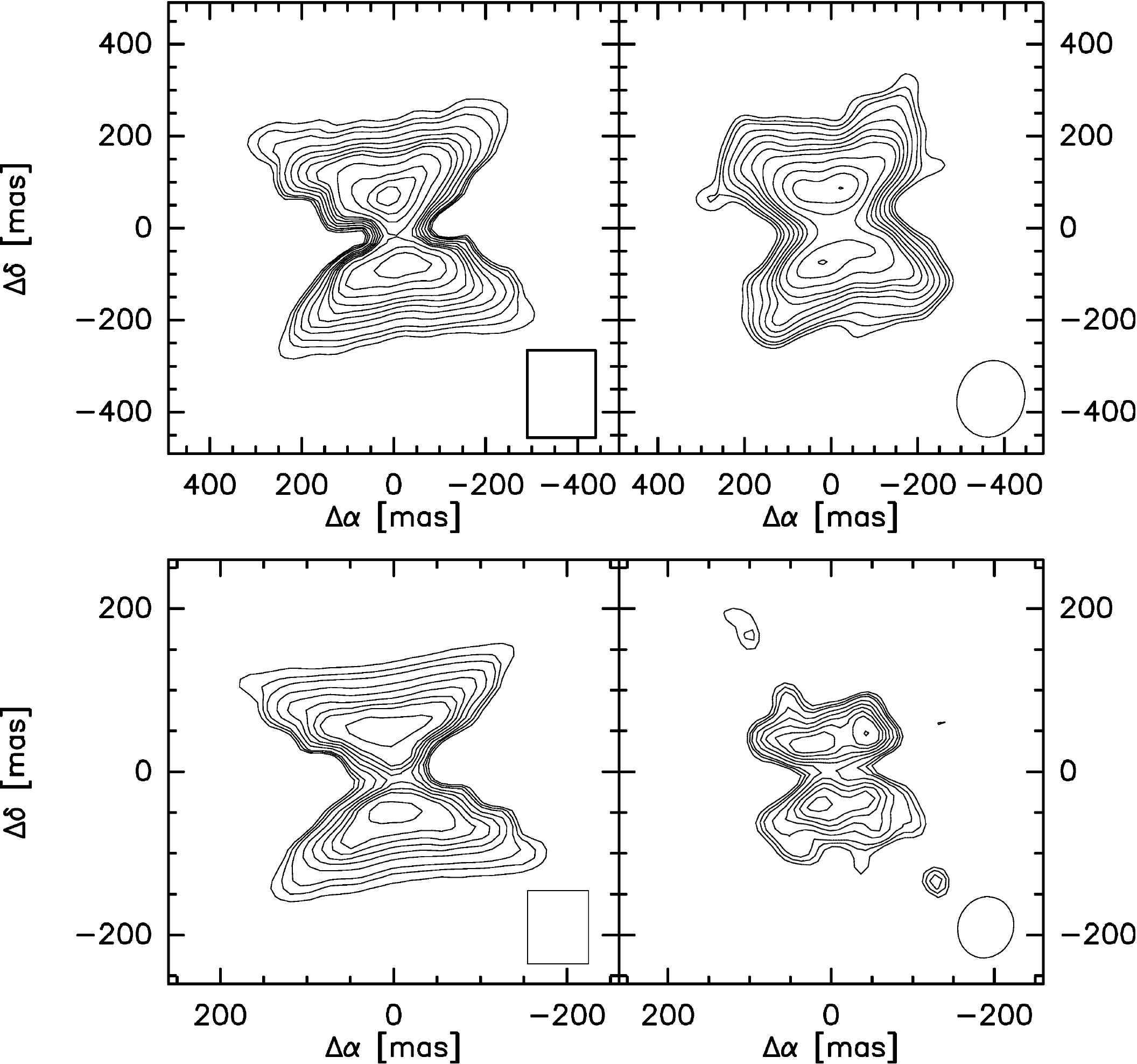}
\caption{Modelled (left) and observed (right panel, by \cite{Tafoya2004}) radio-continuum maps for 1.3 cm (top) and 7 mm (bottom). Contours are -5, -4, 4, 5, 6, 8, 10, 12, 14, 16, 20, 25, 30, 35, 40, 50, 60, 70, 80, 90, 100, and 110 times the rms of the observational VLA image (727 and 859 $\mathrm{\mu Jy}$ beam$^{-1}$ for the 1.3 cm and 7 mm images, respectively). The modelled maps were obtained by the convolution of the original data with a box of the same size as the HPBW of the observations) and rotating the image by 8$\degr$ anticlockwise.}
\label{figure:radio_continuum}
\end{figure}

The radio-continuum maps obtained with the best angular resolution (at 7 mm, see Fig. \ref{figure:radio_continuum}) clearly show a bipolar morphology; there is a prominent dark lane in the East-West direction with a position angle of 8$\degr$ (measured counterclockwise from East). The lack of free-free emission is interpreted as being caused by a neutral circumstellar disk. This interpretation is also supported by the dust emission from the disk in the near-infrared (\cite{Danchi2001}) and its polarization perpendicular to the disk (\cite{Aitken1990}, \cite{Yudin1996}). The radio-continuum emission shows an ionized bipolar structure arising above and below the neutral disk. The geometry of the ionized outflow has been modelled by considering a double-cone. The radio-continuum images are used to constrain the semi-opening angle of the double cone, $\theta_\mathrm{a}$ (Fig. \ref{figure:figura}), to $\sim$ 57$\degr$. This opening angle is consistent with that found with a semi-analytical model of ionized winds around sgB[e] stars (\cite{Kraus2003}). The radio-continuum maps also show that for a given radius, the intensity is larger close to the neutral disk. In the isothermal case, this fact indicates that the electron density decreases from the edge of the ionized disk toward the revolution axis ($0<\theta_0<\infty$).

\begin{figure}
\centering
\includegraphics[width=8cm]{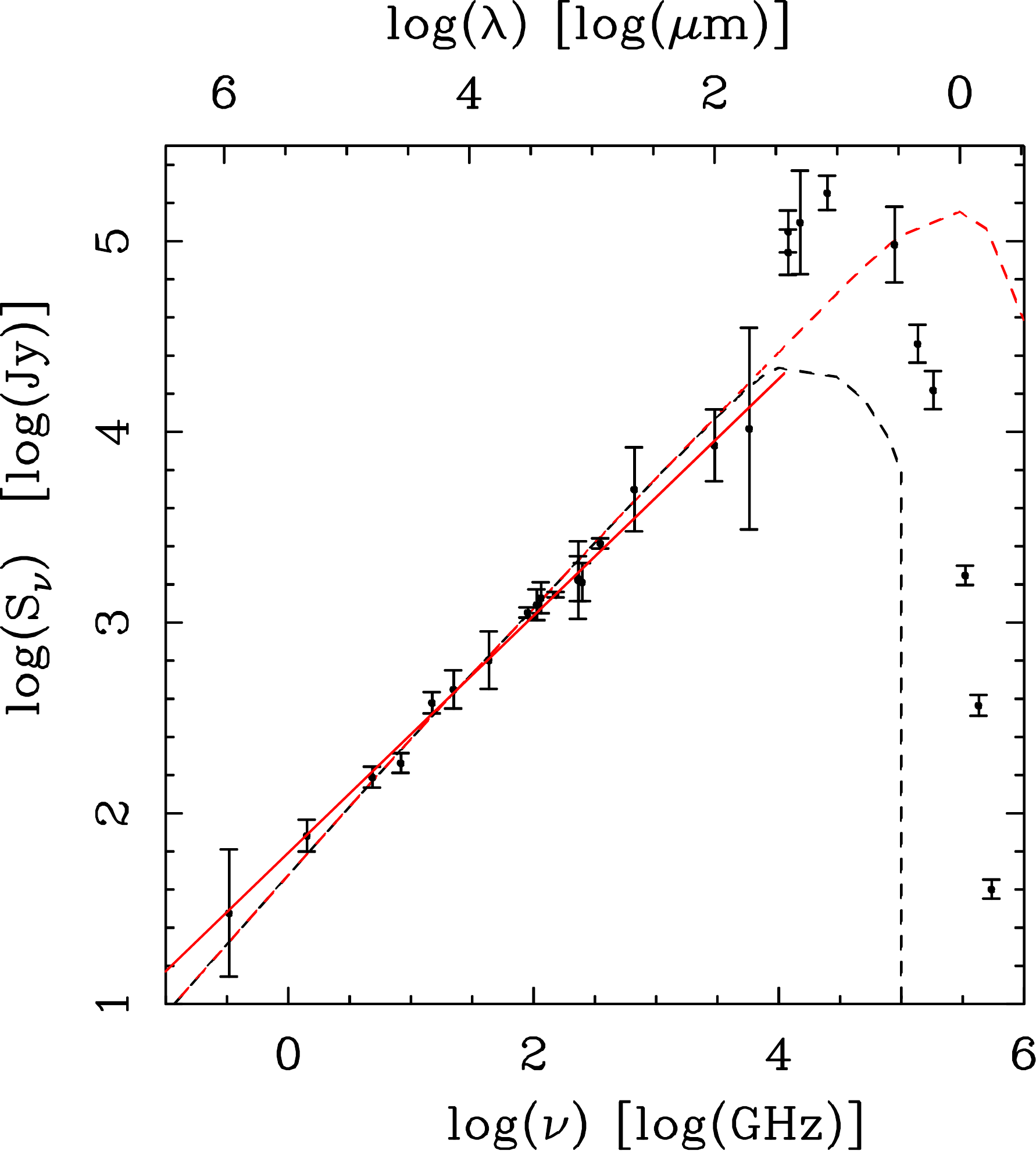}
\caption{Observed (vertical bars) and modelled (dashed lines) SEDs for MWC349A. The black and red dashed lines represent the predictions for ionized stellar winds extending down to $r_{\mathrm{min}}=3$ and 0.05 AU, respectively. The red solid line shows the linear fit to the observational intensities (with $\nu < 5.77$ THz), yielding a spectral index of 0.62. Observational data were obtained from \cite{Allen1973}, \cite{Altenhoff1994}, \cite{Beichman1988}, \cite{Harvey1979}, \cite{Lee1970}, Mart\'in-Pintado et al. (1989a), \cite{Martinpintado1994}, \cite{Planesas1992}, \cite{Sandell2011}, \cite{Schwartz1980}, \cite{SimonDyck1977}, and Tafoya et al. (2004). All the observational data available so far was concisely compiled by \cite{Lugo2004}.}
\label{figure:SED}
\end{figure}

Next we have constrained the physical structure of MWC349A by considering the whole set of continuum measurements available to date for the spectral range covering from radio-frequencies to the mid-IR range, where there is a clear excess of emission with respect to that expected from the free-free emission of the wind ($\nu \sim $ 6000-10000 GHz as shown  in Fig. \ref{figure:SED}). For simplicity we have assumed an isothermal wind expanding at nearly constant velocity in order to explain its radio-continuum spectral index of 0.62 (Fig. \ref{figure:SED}).

The best fits to the radio-continuum images and SED (see Figs. \ref{figure:radio_continuum} and \ref{figure:SED}, respectively) were obtained by considering the values of $n_\mathrm{R}$, $b_\mathrm{d}$, $T_\mathrm{e}$, $N_\mathrm{e}\left(r=1,\theta=\theta_\mathrm{a}\right)$, and $\theta_0$ shown in Table \ref{table:inputs}. The value  derived for $\theta_0$ is consistent with that obtained by White \& Becker (1985). Based on the SED, which is rising until at least 5.8 THz (Fig. \ref{figure:SED}), we derive an inner radius of the ionized wind of $r_{\mathrm{min}}=3$ AU or smaller, since larger $r_{\mathrm{min}}$ would imply optically thin free-free emission at $\nu<5.8$ THz. We note that the quality of the fit is not very sensitive to the consideration of a thin layer of ionized material located next to the neutral disk with different electron temperature, $T_\mathrm{d}$ (e.g. 9450 K obtained by the fit of the RRL profiles). Although the set of input parameters that best fit the SED is not unique, the fit to the radio-continuum maps breaks this degeneracy. Figure \ref{figure:radio_continuum} shows that our radio-continuum predictions are in very good agreement with the maps reported by Tafoya et al. (2004). We also note that the constraints on the parameters that are affected by some uncertainties were improved by using the RRL observations as described below.

\subsection{Constraining the orientation, kinematics, and physical structure of MWC349A on the basis of the H30$\alpha$ centroid map and profile}
\label{section_H30alpha}

Mart\'in-Pintado et al. (2011) explained the H30$\alpha$ RRL emission obtained with the Plateau de Bure Interferometer by considering a model with two kinematic components made up of an expanding outflow and an ionized Keplerian rotating disk (as represented in Fig. \ref{figure:figura}). Besides the explanation of the velocity peak separation of the maser spikes, the biggest achievement of the quoted model was to explain the behaviour of the H30$\alpha$ RRL centroid map. Its interpretation is not straightforward since $\beta_{mn}<0$ holds for a wide range of electron density and electron temperature values (\cite{Strelnitski1996b}) and, therefore, the stimulated amplification is not produced in a small region with very specific physical conditions, but in an extended region (see bottom panels of Fig. \ref{figure:spikes}). Thus, in the case of RRL maser emission, the centroid map must be interpreted as the averaged position where the emission for every radial velocity is produced.



\begin{figure}
\centering
 \resizebox{\hsize}{!}
{\includegraphics[width=0.85\textwidth]{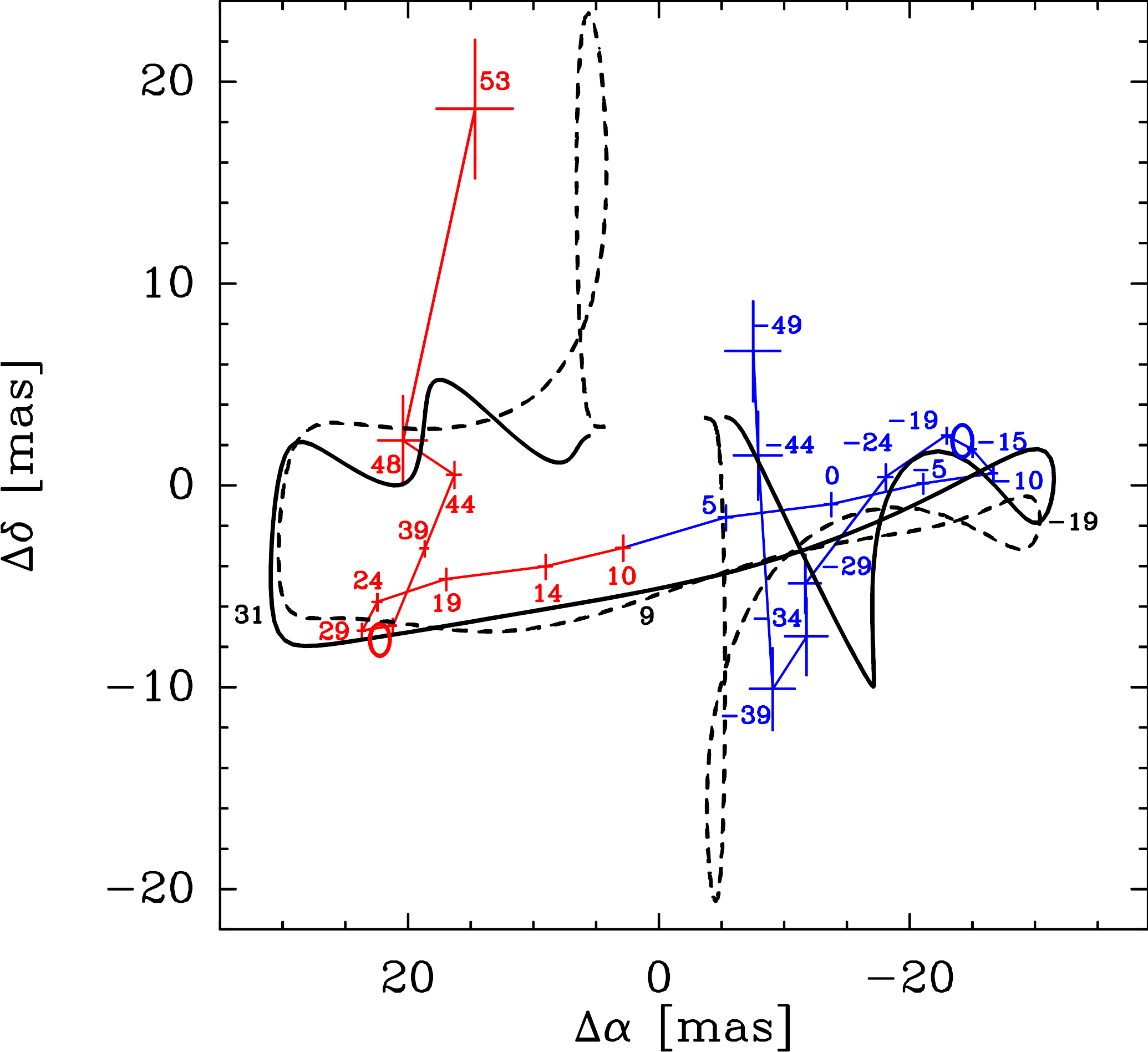}}
 \caption{Model prediction for the relative centroid positions of the H30$\alpha$ emission (solid and dashed thick lines for the cases with rotating and non-rotating outflow) in the velocity range between -64 km s$^{-1}$ and 58 km s$^{-1}$ superimposed on the observed data as blue and red thin lines for MWC349A. The positions of the line velocity channels are shown as crosses labeled by their LSR velocities. The blueshifted and redshifted maser spikes (-16 and 32 km s$^{-1}$, respectively) peak at the blue and red circles. The associated errors are shown by horizontal and vertical bars. The offsets are in milliarcseconds (mas) relative to the continuum centroid ($\alpha_{\mathrm{J2000}}=20^{\mathrm{h}}32^{\mathrm{m}}45^{\mathrm{s}}.54$, $\delta_{\mathrm{J2000}}=40\degr39'36.8''$).}
\label{figure:centroid_map}
\end{figure}
\begin{figure}
\centering
\includegraphics[width=9cm]{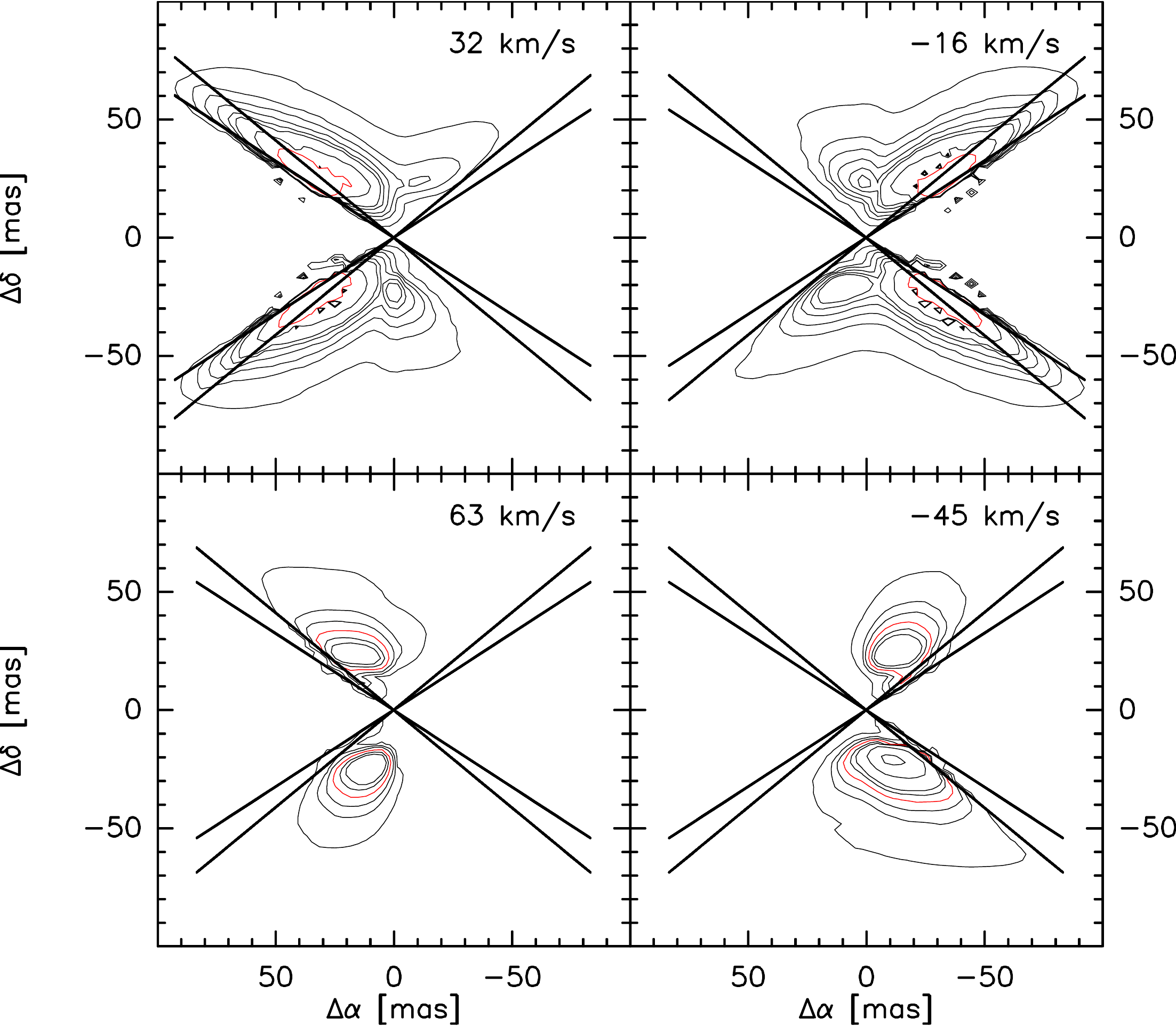}
\caption{Predicted intensity line emission ($\Delta \mathrm{v}_\mathrm{r}=1$ km s$^{-1}$) of the H30$\alpha$ at radial velocities of 32 (top left), -16 (top right), 63 (bottom left), and -45 km s$^{-1}$ (bottom right). Contour levels are 0.5, 1.5, 2.5, 4.0, 6.0, 8.0, 20.0 and 80.0 mJy. The contour level of 80.0 mJy (corresponding to an optical depth of -3.6) of the upper panels (red contour levels) contain $\sim$ 80\% of the total emission, while the contour level of 4.0 mJy (corresponding to an optical depth of -0.8) of the bottom panels (red contour levels) contain $\sim$ 50 and 70 \% of the total emission (left and right, respectively). It shows that at the radial velocities of the spikes (upper panels), the emission mainly originates at points located close to the projection of the ionized disk (whose edges are represented as straight lines for $\theta_i=0$) into the source's plane of symmetry perpendicular to the line-of-sight.}
\label{figure:spikes}
\end{figure}

Mart\'in-Pintado et al. (2011) have shown that the centroid map of the H30$\alpha$ (see Fig. \ref{figure:centroid_map}) for radial velocities comprised between the two maser spikes (at -16 km s$^{-1}$ and 32  km s$^{-1}$, respectively) is well fitted to a straight line. This behaviour suggests that the emission between these radial velocities mainly arises from the Keplerian ionized layer. This is also supported by the spatial distribution of the emission predicted by our model (see Fig. \ref{figure:spikes}). 

The physical reason for the emission from the maser spikes to originate mainly in the ionized Keplerian disk is because the electron densities in this  region are close to that optimum for maser amplification of the H30$\alpha$ RRL. For example, an electron density of 4.0$\times$10$^7\ \mathrm{cm}^{-3}$ (the optimum density for an electron temperature of $10^4\ \mathrm{K}$; \cite{Strelnitski1996b}) is reached at the border between the ionized and the neutral disk at a radius $\sim$50 AU according to the results of our model.

On the other hand, the emission at larger radial velocities mainly arises from a region farther away from the star and located outside of the ionized circumstellar disk. Although the overall H30$\alpha$ centroid map clearly constrains the kinematic features of the outflow (\cite{Martinpintado2011}), it leaves open some uncertainties due to the high number of input parameters required for the modelling. However, the RRL profiles lead to strong constraints on their values. 

\begin{figure}
\centering
\includegraphics[width=8cm]{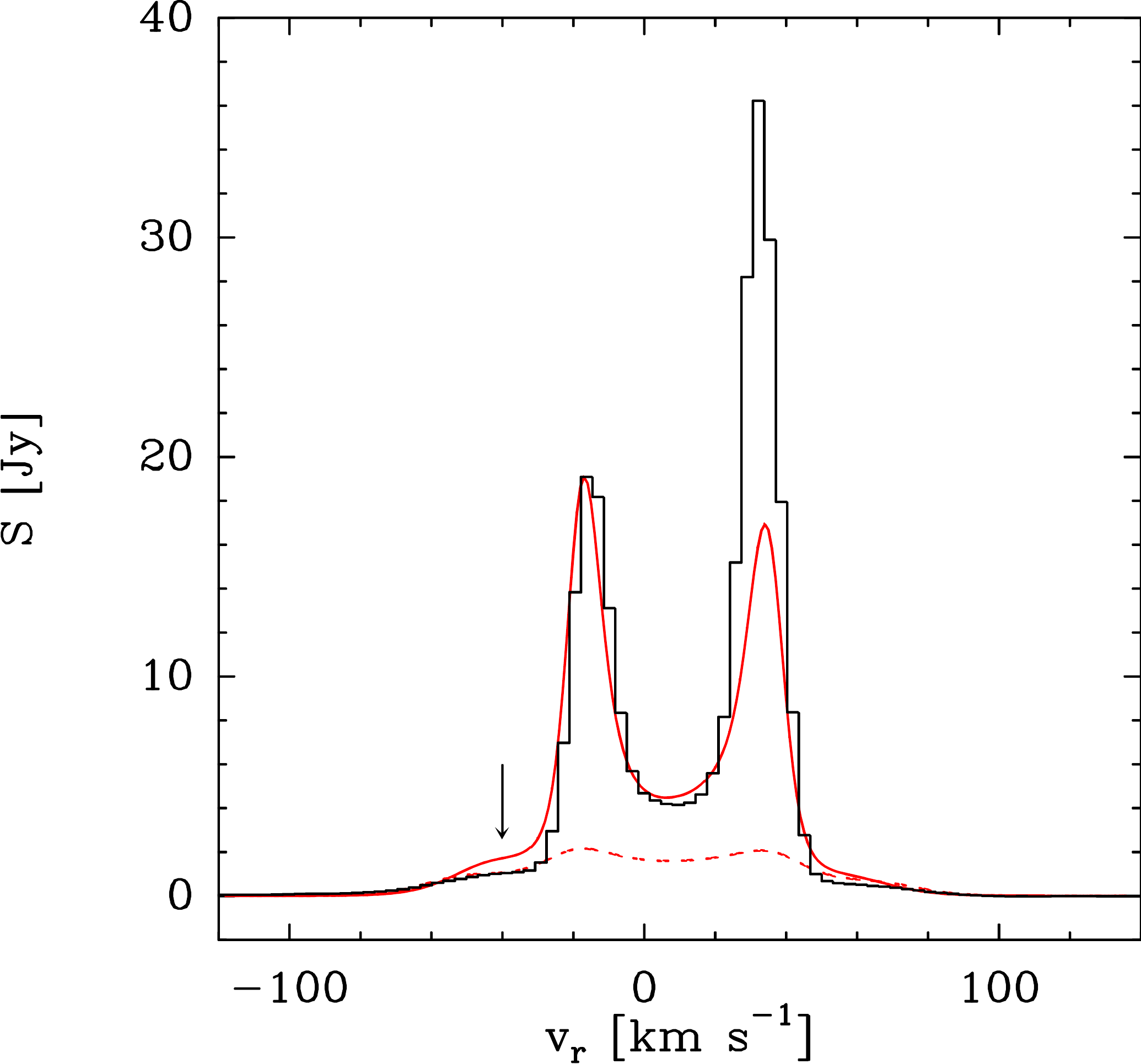}
\caption{Observed (histogram) and modelled profiles (lines) of the H30$\alpha$. The red dashed line is the best profile obtained by using the Walmsley departure coefficients (with the input parameter values as given in \cite{Martinpintado2011}), while the solid red line is the modelled profile obtained with the set of input parameters shown in Table \ref{table:inputs} and by using the Storey \& Hummer $b_n$ coefficients. The derived amplification assuming the Walmsley coefficients is lower because of their lower negative $\beta_{mn}$ values (see Fig. 2). The arrow indicates the wing hump at the blue-shifted velocities.}
\label{figure:H30alpha_profiles}
\end{figure}

In particular, the H30$\alpha$ RRL profile shows two different spectral features (see Fig. \ref{figure:H30alpha_profiles}), a narrow double-peak maser emission at radial velocities of $\sim$ -16 and 32 km s$^{-1}$ arising from the ionized Keplerian disk, and two broader features at slightly larger radial velocities, $\sim-45$ and $\sim71$ km s$^{-1}$, showing a clear asymmetry (hereafter wing humps).  Although the two components could be confused, the model shows that while the features of the narrow maser spikes depend on the kinematics of the Keplerian layer, the 
 wing humps depend on the characteristics of the outflow. As described in Sect. \ref{dependence_on_inputs}, these two maser components arising from two different regions with different kinematics were used to constrain some of the input parameters of the model.                         




Figure \ref{figure:H30alpha_profiles} shows that while the computed profile obtained by using the set of $b_n$ coefficients provided by \cite{Walmsley1990} reproduces the main features of the observational profile, it clearly fails to predict its intensity. Neither set of parameters (with an electron temperature larger than 4000 K) can explain the intensities of the observed spikes. On the contrary, the \cite{Storey1995} departure coefficients with the input parameters given in Table \ref{table:inputs} are able to reproduce much better both the features of the profile and its peak intensity. Thus we have assumed the \cite{Storey1995} departure coefficients in the model in order to significantly improve the fit of the maser spike intensities. Unlike the profile, the modelled H30$\alpha$ centroid map remains similar for both set of coefficients.

We note that the observational profile of both the H30$\alpha$ and the rest of double-peaked lines (see Sect. \ref{section_profiles}) show significant differences between the intensities of the blue- and red-shifted peaks, while our model predicts roughly the same intensities for both spikes. However, as we can see in Fig. \ref{figure:H30alpha_profiles}, the maser peak intensities are strongly dependent on the set of departure coefficients assumed by MORELI. On the other hand, since most of the emission of the maser spikes comes from very small regions (see Fig. \ref{figure:spikes}) and it is strongly dependent on optical depth, clumps with slightly different electron temperatures or electron densities would produce significant changes in the peak intensities. Thus, differences in the electron temperature and electron density of clumps located in the blue- and red-shifted regions of MWC349A, together with the uncertainties in the departure coefficient values (see Sects. \ref{departure_coefficients} and Fig. \ref{beta_coefficients}), could explain the observed asymmetries between the two maser peaks. Likewise, the presence of different clumps with slightly different physical conditions over time would also explain the rapid time variations of maser RRLs (\cite{Martinpintado1989b}). This variability is seen in time scales as short as 30 days (\cite{Thum1992}).

To explain the H30$\alpha$ maser peak-to-peak separation (see Table \ref{table:peak_separation}), we have considered a central mass for MWC349A of 38 $\mathrm{M}_{\odot}$, unlike the $\sim$ 60 $\mathrm{M}_{\odot}$ obtained if we use the \cite{Walmsley1990} $b_n$ coefficients (\cite{Martinpintado2011}). As thoroughly discussed by \cite{Kraus2009}, the location of MWC349A in the HR diagram is subjected to a high uncertainty, particularly because of poor determination of its luminosity ($\log{T_\mathrm{eff}}=$4.37$\pm$0.07, $\log{\left(L/L_{\odot}\right)}=$5.7$\pm$1.0). This makes its location in the HR diagram consistent with the evolutionary tracks of a wide range of central masses of rotating and non-rotating massive stars (\cite{Schaller1992} and \cite{Meynet1994}). Nevertheless, if we take into account just the estimated central values of the effective temperature and luminosity ($\log{T_\mathrm{eff}}=$4.37, $\log{\frac{L}{L_{\odot}}}=$5.7), we derive a mass for the star of $\sim$ 38 $\mathrm{M}_{\odot}$. This value is consistent with our results. These results seem to suggest again that the Storey \& Hummer departure coefficients provide a better description of the stimulated amplification of the radiation in MWC349A.

For a 38 $\mathrm{M}_{\odot}$ central mass, the best fit for both the H30$\alpha$ centroid map (Fig. \ref{figure:centroid_map}) and the RRL profiles was obtained by considering the following input parameter values: i) outflow terminal and turbulent velocities of $\sim$60 and 15 km s$^{-1}$, respectively; ii) Keplerian rotation velocity component for the the outflow (see Fig. \ref{figure:centroid_map}) in the same sense as the ionized disk; iii) a Keplerian rotating ionized disk layer with an opening angle of 6.5$\degr$ located next to the neutral disk and extended from 0.05 AU to 130 AU; iv) an electron temperature of 9450 K for the ionized Keplerian disk; and v) inclination angle of the disk axis with respect to the plane of the sky of 8$\degr$, where the front-side is tipped up. Next, we discuss how changes in the value of these parameters affect the fitting of the observational data.

\subsubsection{Dependence of the modelled H30$\alpha$ centroid map and RRL profiles on the input parameters}
\label{dependence_on_inputs}

\paragraph{Inclination angle, $\theta_\mathrm{i}$}
$ $ \\ \\ The inclination angle of the disk axis is the key parameter that makes the emission asymmetric with respect to the east-west plane. Therefore, it is responsible for the north-south loops of the H30$\alpha$ centroid map shown in Fig. \ref{figure:centroid_map}. The larger the inclination angle, the larger the asymmetry, with larger heights for the loop peaks at high radial velocities ($\sim$ 50 km s$^{-1}$), which correspond to the emission of the outflow. Thus, the inclination angle is clearly constrained between 4.5$\degr$ and 15$\degr$, with the front side of the disk tipped up with respect to the line of sight. Thus, we explain that the red-shifted and blue-shifted loops (see Fig. \ref{figure:centroid_map}) occur south and north of the disk respectively.

\begin{figure*}
\centering
\includegraphics[width=16cm]{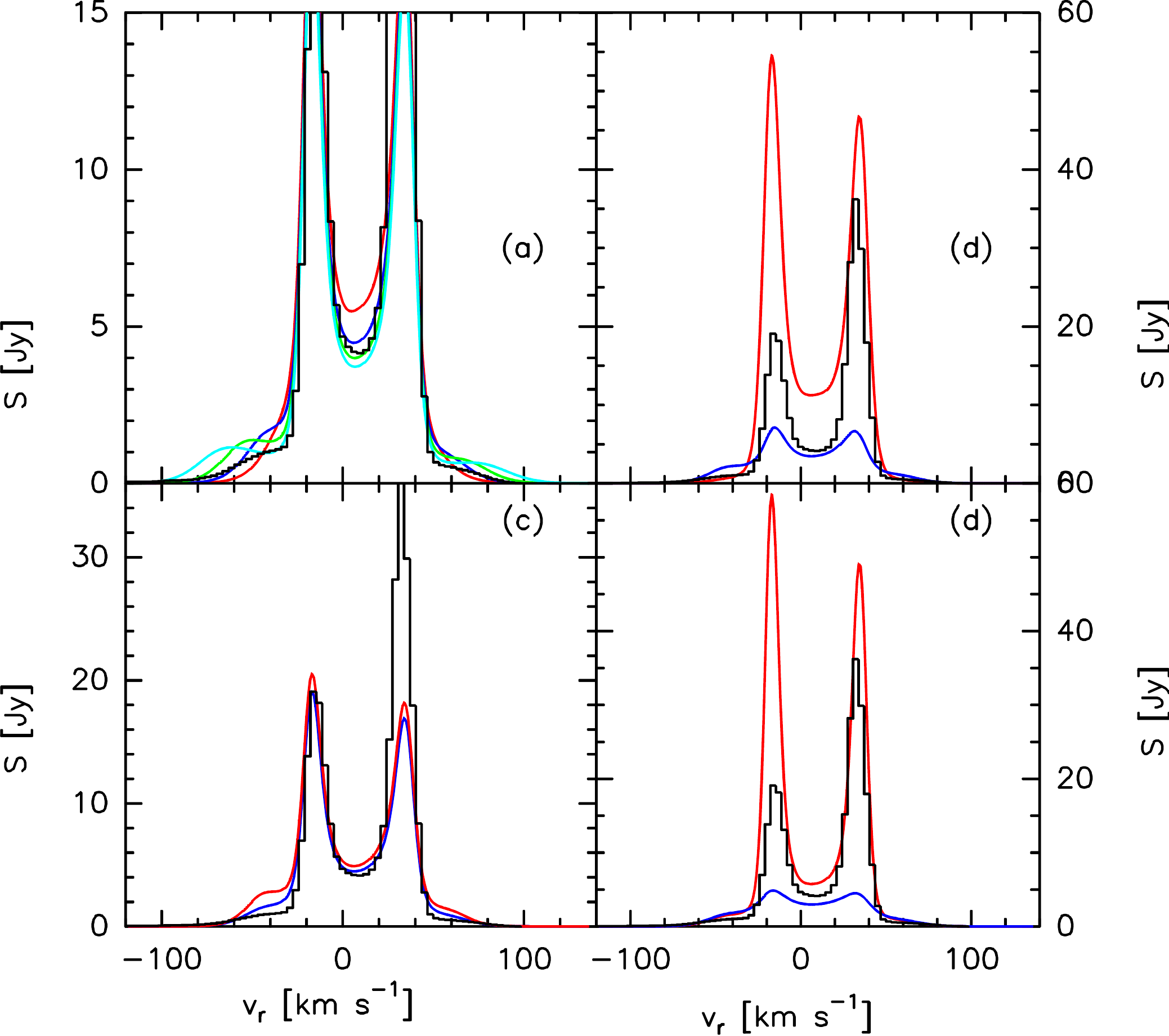}
\caption{Observational (black histogram) and predicted H30$\alpha$ profiles varying different input parameters: \\
(a) predicted H30$\alpha$ profiles for terminal velocities of the ionized wind of 40, 60, 80, and 100 km s$^{-1}$ (red, blue, green, and cyan). It is clearly shown that the wing humps are shifted to larger radial-velocities with increasing terminal velocities.\\
(b) predicted H30$\alpha$ profiles for opening angles for the Keplerian ionized layer, $\theta_\mathrm{d}$, of 4.5$\degr$ and 15$\degr$ (blue and red lines) \\
(c) predicted H30$\alpha$ profiles for electron temperatures for the ionized outflow, $T_\mathrm{o}$, of 10000 and 12000 K (red and blue lines).\\
(d) predicted H30$\alpha$ profiles for electron temperatures for the Keplerian ionized layer, $T_\mathrm{d}$, of 8500 and 12000 K (red and blue lines).
} 
\label{figure:H30alpha_profiles_with_diff_inputs}
\end{figure*}

\paragraph{Terminal velocity, v$_0$}
$ $ \\ \\ The model also shows that the radial velocity range where the maser emission of the outflow is produced depends on its terminal velocity. Thus, we can rule out terminal velocities larger than 100 km s$^{-1}$ since then the peaks of the wing humps would be located at larger radial velocities, clearly distinguished as high-velocity shoulders (see Fig. \ref{figure:H30alpha_profiles_with_diff_inputs}a). In contrast, a too low terminal velocity would imply that the wing hump emission would appear at nearly the same radial velocities as the maser spikes and, therefore, we would not be able to distinguish the contribution of the two different kinematic components (ionized disk and outflow). Thus, we conclude that the terminal velocity must be comprised between 40 and 100 km s$^{-1}$. This is also supported by the way in which the predicted H30$\alpha$ centroid map changes with this velocity. However, the fit to the linewidths of the H41$\alpha$ and H76$\alpha$ RRLs (cf. Sect. 5.1) provided the strongest constraint on the outflow terminal velocity. Thus, finally we considered a value of 60 km s$^{-1}$. 



\paragraph{Turbulent velocity, $\mathrm{v}_\mathrm{tu}$}
$ $ \\ \\ Turbulent motions in the ionized wind are needed in order to explain the lack of a double-peaked profile observed for the H66$\alpha$ line profile (see Sect. \ref{section_profiles}). We note that the two observed peaks would be due to stimulated emission from the ionized outflow and not from the disk, opposite to that found in mm RRLs. Thus, we need to consider a turbulent velocity. We conclude that the best choice for the turbulent velocity to explain the line profiles and the centroid map is $v_\mathrm{tu}\sim$15 km s$^{-1}$, which is comparable to the sound speed for 12000 K. Larger turbulent velocities would increase the predicted linewidths.







\paragraph{Opening angle of the Keplerian ionized layer, $\theta_\mathrm{d}$}
$ $ \\ \\ Given a semi-opening angle for the ionized wind ($\theta_\mathrm{a}\sim$57$\degr$ as constrained from radio-continuum maps, Sect. \ref{section_H30alpha}), the contribution of the emission arising within the ionized Keplerian ionized disk is determined by the opening angle of this disk, $\theta_{\mathrm{d}}=\theta_\mathrm{a}-\theta_\mathrm{w}$ (see Sect. \ref{kinematics_modelling}). The larger the opening angle of the disk, the lower the flux arising from the outflow and, therefore, the weaker the wing hump intensities. Thus, too low opening angles for the disk would significantly increase the wing hump intensities compared to observed levels. In addition, in this case the asymmetry between the red- and blue-shifted wing humps would also increase. We conclude that our model is very dependent on $\theta_\mathrm{d}$ for angles smaller than 6$\degr$. In particular, we rule out that $\theta_\mathrm{d}$ could be smaller than 4.5$\degr$  (see Fig. \ref{figure:H30alpha_profiles_with_diff_inputs}b). Likewise, if one considers a larger angle (i.e. 15$\degr$), the wing hump intensities would decrease and the maser peak intensities would increase above the observed values (see Fig. \ref{figure:H30alpha_profiles_with_diff_inputs}b). Taking into account the aforementioned considerations, we conclude that the opening angle of the Keplerian ionized layer must be between 4.5$\degr$ and 15$\degr$.

\paragraph{Electron temperatures}
$ $ \\ \\ Regarding the electron temperature of the ionized outflow and of the ionized Keplerian disk, our model shows that the centroid map is less sensitive to changes in the electron temperature than the line profiles. We consider an electron temperature of 12000 K for the outflow to adequately fit the radio-continuum data (Sect. 2) and the intensities of the wing humps. If we considered lower values, the maser effect would increase and, consequently, also the wing hump intensities as shown in Fig. \ref{figure:H30alpha_profiles_with_diff_inputs}c.


On the other hand, we have considered a different electron temperature for the Keplerian ionized disk, $T_\mathrm{d}$, from that of the ionized outflow to increase the intensity of the maser spikes. Since the maser intensities depend strongly on the temperature (as shown in Fig. \ref{figure:H30alpha_profiles_with_diff_inputs}d), we constrain the electron temperature in the Keplerian ionized disk between 8500 and 11000 K. We adopt $T_\mathrm{d}=$ 9450 K as the final value for the Keplerian ionized layer.





\paragraph{Size of the disk, $r_d$}



$ $ \\ \\ In the equatorial region located beyond the circumstellar disk, the observed radio-continuum emission at cm wavelengths (\cite{Tafoya2004}) must be due to the ionization of the gas by the dust-scattered ionizing radiation. This is because we observe emission in the equatorial region for radii larger than that of the beam for the 2, 3.6, 6, and 20 cm radio-continuum images (\cite{White1985}, Rodr\'iguez \& Bastian 1994, Cohen et al. 1985, and \cite{Tafoya2004}). From the beam sizes and the constrained semi-opening angle for the ionized gas, $\theta_{\mathrm{a}}$, we derive an upper limit to the radius of $\sim$ 130 AU. This upper limit is consistent with the derived size of 75 AU from the 3.8 $\mathrm{\mu m}$ image (Danchi et al. 2001). On the other hand, since the Keplerian rotating ionized disk is formed by the photoevaporation of the neutral disk, we assume that it reaches at least the same radius as the neutral disk. Thus, we have assumed a radius for the Keplerian ionized disk of 130 AU. We have to stress that the results of the model are only very sensitive to the disk size for a radius smaller than 60 AU (50 mas) since the bulk of the mm-wavelengths RRL emission (for which the maser emission is mainly produced in the ionized disk) comes from a region within this radius (i.e. as happens in the H30$\alpha$ line seen in Fig. \ref{figure:spikes}). 







\subsection{Fit to the line-integrated intensities}
\label{section_int_line_int}

\begin{figure}
\centering
\includegraphics[width=9cm]{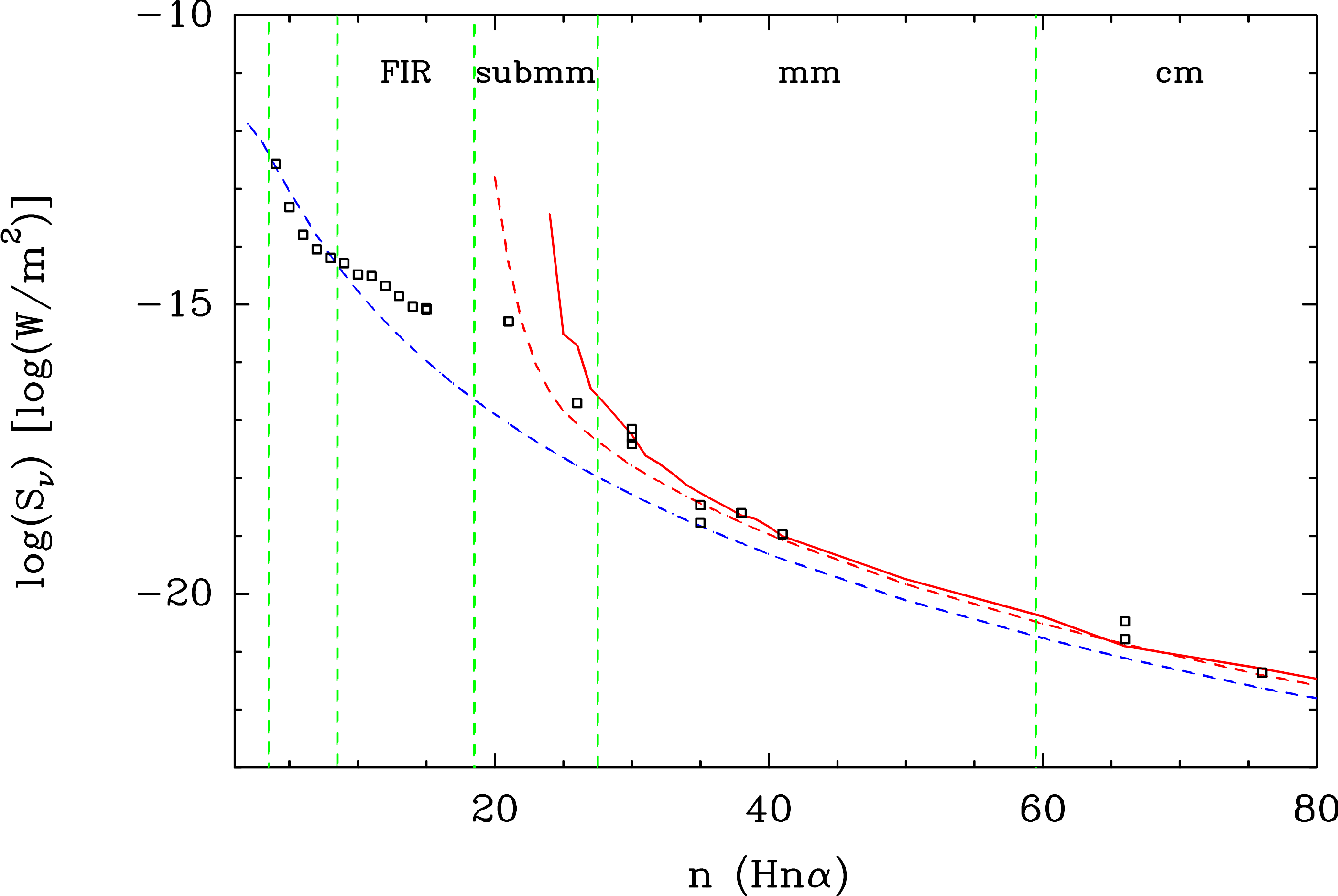}
\caption{Integrated line intensities for the whole range of quantum numbers $n$ of the observed Hn$\alpha$ RRLs. Open squares refer to the observational data, while the lines show the LTE (blue line) and non-LTE prediction of our model (dashed and solid red lines for the cases of the Walmsley and the Storey \& Hummer $b_n$ coefficients respectively). With green dashed lines we show the limits among the different wavelength ranges. References for the observational data: H4$\alpha$ to H15$\alpha$ Thum et al. (1998); H21$\alpha$ \cite{Thum1994b}; H26$\alpha$ and H30$\alpha$ Thum et al. (1994a); H30$\alpha$ Mart\'in-Pintado et al. (1989a); H35$\alpha$ and H38$\alpha$ \cite{Martinpintado1994}; H41$\alpha$ Mart\'in-Pintado et al. (1989a); H66$\alpha$ Loinard \& Rodr\'iguez (2010); H66$\alpha$ Mart\'in-Pintado et al. (1993); H76$\alpha$ Escalante et al. (1989).}
\label{figure:int_line_fluxes}
\end{figure}

Using the structure and physical properties derived in previous sections, we carry out the modelling of the integrated intensity of RRLs to study the frequency range where the maser emission is taking place. For that purpose, we predict the Hn$\alpha$ integrated line intensities both for the non-LTE and LTE cases for the quantum numbers from n=4 to n=80 as shown in Fig. \ref{figure:int_line_fluxes}. For the non-LTE case, we show the predictions using both the Walmsley (1990) and the Storey \& Hummer (1995) departure coefficients. 

Firstly, we see that the two integrated line flux predictions obtained under the non-LTE assumption for the two sets of departure coefficients converge for Hn$\alpha$ lines at low frequencies (with n$\gtrsim$41). The non-LTE predictions agree extremely well with the observations and show a constant offset with respect to the LTE prediction. This is a clear sign that RRL emission is stimulated at low-frequencies as claimed by Mart\'in-Pintado et al. (1993) and as supported by the analysis of RRLs profiles described in Sect. \ref{section_profiles}.

For RRLs with quantum numbers smaller than 41, we can see that the predicted integrated intensities for the non-LTE case become much larger than the LTE prediction, especially for RRLs with $n<30$. This is what one expects since the lower the electronic level, $n$, the larger the importance of the mechanisms causing the electron population inversion, as in agreement with the departure coefficients (Strelnitski et al. 1996b). Only the non-LTE case is able to explain the large increase of the observed integrated line intensities. The large dependence of the integrated line intensity on the quantum numbers for $21\leq n\leq41$ is proof that the dominant process of amplification of the emission at such RRLs is maser.


Likewise, it is remarkable that the predictions obtained with the Storey \& Hummer (1995) coefficients are different from those derived using the Walmsley (1990) coefficients. This allows us to discriminate which set of departure coefficients is the best to describe the maser amplification under the physical conditions of MWC349A. For RRLs with $n\geq30$, we clearly see that the Storey \& Hummer (1995) departure coefficients are more appropiate since their integrated line intensities agree extremely well with the published data for $30\le n<41$.

Thus, we conclude that the maser emission at such frequencies is consistent with the physical conditions and the kinematics that we have deduced based on the radio-continuum and RRLs profiles. This supports the proposed kinematic model and the estimated physical parameters. Nevertheless, the integrated line intensity predictions increase very strongly with the frequency compared to that observed for $n<30$, as one expects from an exponential increase of the brightness temperature with the optical depth (Sect. \ref{saturation_section}). It indicates that saturation effects in the masers play an important role for $n<30$. Assuming an equivalent maser beam solid angle, $4\pi / \Omega_{\mathrm{m}}$, of $60$ (\cite{Thum1994a}), we have checked that the degree of saturation, $J_{\nu \mathrm{, sat}}/J_{\nu}$, strongly increases from the H30$\alpha$ to the H26$\alpha$, and especially for the H21$\alpha$. Table \ref{table:degree_sat} displays the radius, electron density, temperature of saturation, brightness temperature and degree of saturation derived from MORELI in those pixels where the upper limit for the degree of saturation is maximum. We clearly see that the degree of saturation for the H26$\alpha$ and, above all, for the H21$\alpha$ are large enough to significantly contribute to the decrease of the maser amplification of the radiation in at least some of the pixels. 






\begin{table}
\caption{Magnitude values in the pixels with maximum degree of saturation.}
\label{table:degree_sat}
\begin{tabular}{cccccc} \hline
Line&$R_{\mathrm{sat}}$ $[\mathrm{AU} ]$ $^{\mathrm{ a }}$& $N_\mathrm{e}$ $[ \mathrm{cm}^{-3}  ]$ & $T_\mathrm{sat}$ $[ \mathrm{K} ]$ $^{\mathrm{b}}$  & $T_\mathrm{B}$ $ [ \mathrm{K} ]$& $J_{\nu \mathrm{, sat}}/J_{\nu}$ $^{\mathrm{b}}$ \\ \hline \hline
H30$\alpha$&66&$2.8x10^7$&$6.5x10^5$&$4.6x10^6$&7 \\ 
H26$\alpha$&49&$5.3x10^7$&$2.1x10^5$&$4.0x10^8$&$1.9x10^{3}$ \\ 
H21$\alpha$&23&$28x10^7$&$2.5x10^5$&$4x10^{20}$&$1x10^{15}$ \\ \hline
\end{tabular}

\begin{list}{}{}
\item[$^{\mathrm{a}}$] Radius of the pixel where $J_{\nu \mathrm{, sat}}/J_{\nu}$ is maximum.
\item[$^{\mathrm{b}}$] Estimated by assuming $4\pi / \Omega_{\mathrm{m}}=60$
\end{list}
\end{table}


Finally, the model shows that the non-LTE effects dominate the emission of RRLs down to $n=7$, while RRLs with smaller quantum numbers seems to be formed under LTE conditions as discussed by Thum et al. (1998).


\section{Analysis for RRL profiles toward MWC349A}
\label{section_profiles}


In the following we will use the inferred kinematic properties of the ionized wind and disk for MWC349A to show that the predicted results of our model are also consistent with the observation of RRLs other than H30$\alpha$. First of all, we will focus in the RRLs with quantum numbers $n>30$. Since these RRLs trace outer regions characterized by gas with smaller electron density than those traced by the H30$\alpha$ RRL, their maser emission is expected to be weaker (Mart\'in-Pintado et al. 1989a). Thus, the modelling of their profiles is less sensitive to the set of departure coefficients considered and to small variations of the input parameters. Furthermore we will also study the profiles of RRLs with $n<30$, which trace the kinematics of the innermost regions of the disk. 

\begin{figure}
\centering
\includegraphics[width=9cm]{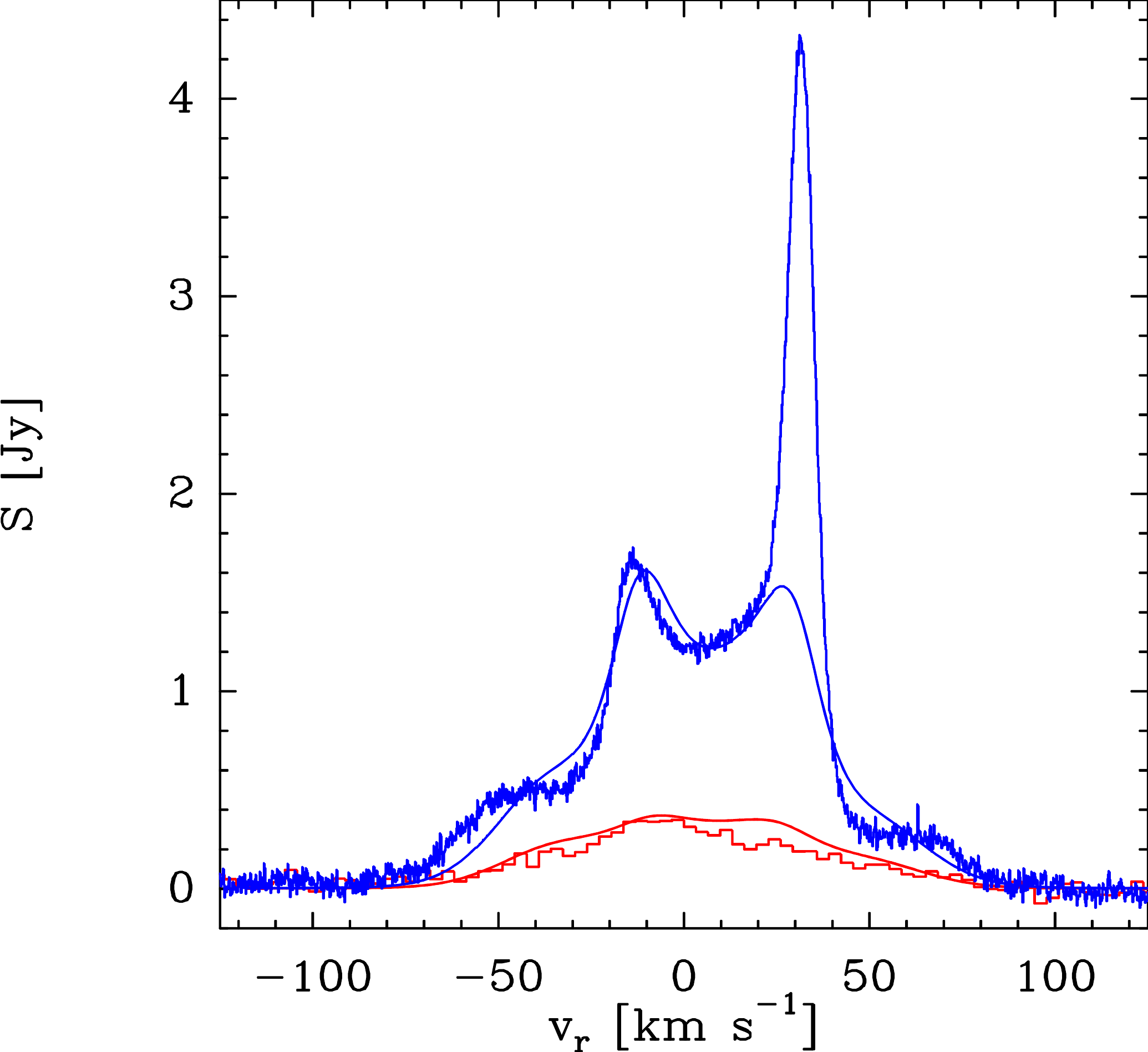} 
\caption{Observed (histogram, unpublished data from C. Thum) and predicted (solid line) RRL profiles for the H35$\alpha$ and H41$\alpha$ shown in blue and red, respectively.}
\label{figure:H35_36alpha_profile_for_letter}
\end{figure}

The predicted profiles should explain the change from the double-peaked profile observed for the H35$\alpha$ (as for the H30$\alpha$) into an apparently simple component for the H41$\alpha$ and also the large change in the peak intensities of these RRLs. This behaviour and the linewidths are perfectly predicted by the model as shown in Fig. \ref{figure:H35_36alpha_profile_for_letter} and in Tables \ref{table:peak_intensities} and \ref{table:peak_separation}. The observed change in the RRL profiles is expected since RRLs at lower frequencies have larger continuum optical depths (see Eq. \ref{continuum_optical_depth}) and trace outer regions where the electron density in the Keplerian ionized layer is lower than the optimum values for maser amplification (i.e. 6.3$\times$10$^6$ cm$^{−3}$ for H40$\alpha$ and 1.4$\times$10$^7$ cm$^{−3}$ for H35$\alpha$; Table 1 in Strelnitski et al. 1996b). For this reason, the H41$\alpha$ emission is mainly dominated by the outflow with little contribution from the disk and, therefore, its profile shows just a single peak. 

\begin{table}
\caption{Comparison between the observed and predicted peak intensities for single-peaked Hn$\alpha$ RRLs toward MWC349A.}
\label{table:peak_intensities}
\centering
\begin{tabular}{cccc}\hline
Line& $\nu \left [\mathrm{GHz} \right ]$& $S_\mathrm{peak, obs} \left [ \mathrm{mJy} \right ]$ & $S_\mathrm{peak, pred} \left [\mathrm{mJy} \right ]$ \\ \hline \hline
H76$\alpha$&14.69&11.7$^{\mathrm{a}}$ & 9.6 \\ 
H66$\alpha$&22.36&24.9$\pm$0.3$^{\mathrm{b}}$ & 16.3 \\ 
H41$\alpha$&92.03&380$\pm$40$^{\mathrm{c}}$ & 370 \\ 
H40$\alpha$&99.02&610$^{\mathrm{d}}$&530  \\ \hline
\end{tabular}

\begin{list}{}{}
\item[$^{\mathrm{a}}$] Escalante et al. (1989).
\item[$^{\mathrm{b}}$] Loinard \& Rodr\'iguez (2010).
\item[$^{\mathrm{c}}$] Mart\'in-Pintado et al. (1989a).
\item[$^{\mathrm{d}}$] \cite{Thum1995}.
\end{list}

\end{table}

We stress that under LTE conditions, we clearly underestimate the peak intensities for all RRLs by factors of two for the H76$\alpha$ and H66$\alpha$ lines, and by larger factors for the rest of the RRLs for which the stimulated emission is stronger. Thus, we rule out the possibility that even apparently single-peaked profiles like that of the H41$\alpha$ line could be formed in LTE.

\subsection{Low-frequency radio recombination lines. The outflow}

\begin{figure}
\centering
\includegraphics[width=8cm]{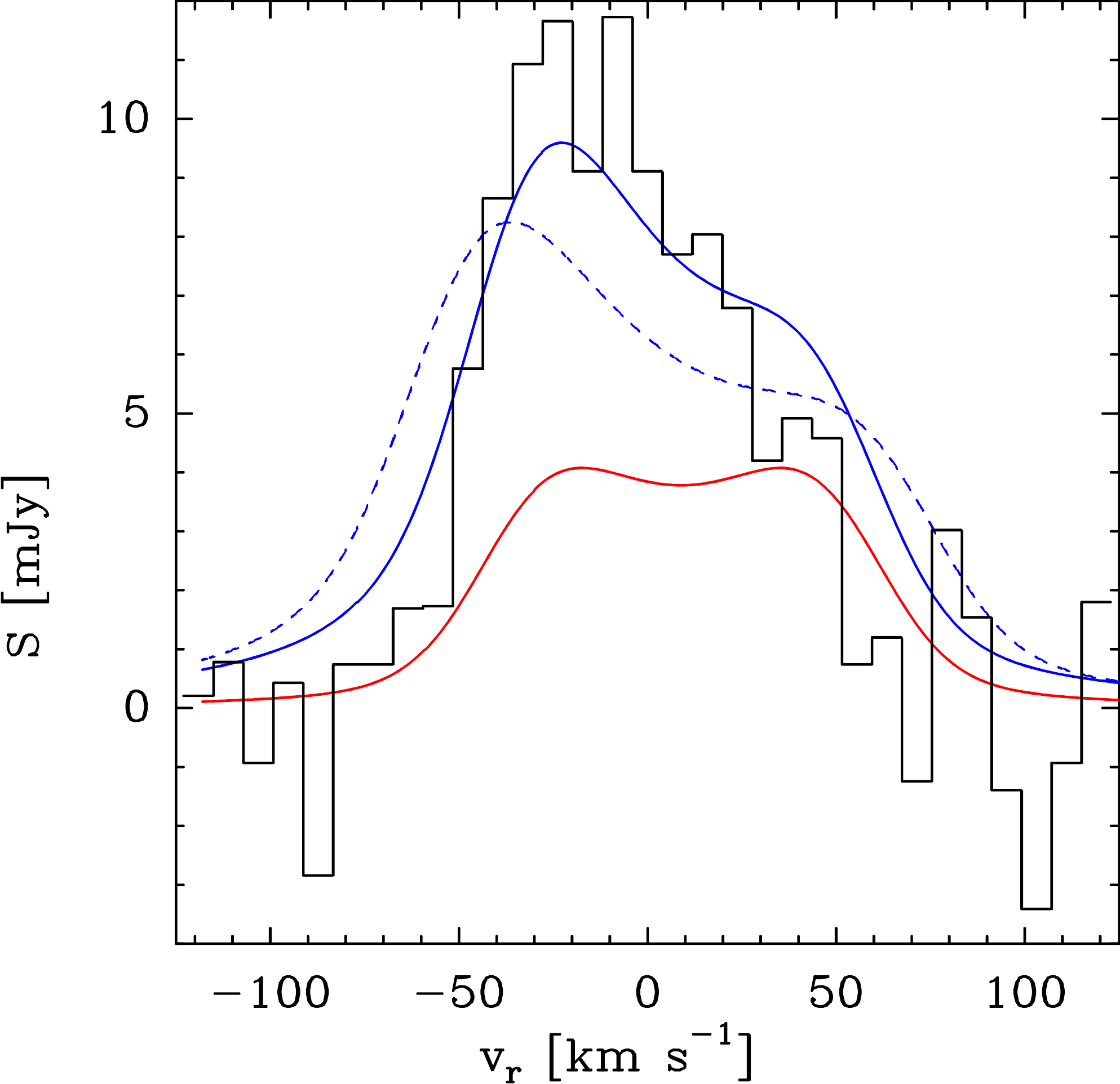}
\caption{Observed (black histogram, \cite{Escalante1989}) and predicted H76$\alpha$ profile in the non-LTE (blue lines) and LTE case (red line). For the non-LTE case, we have plotted the predicted profiles for two different outflow terminal velocities, 60 and 80 km s$^{-1}$ (solid and dashed lines, respectively). We clearly see that the line width is only explained if one assumes $60$ km s$^{-1}$.}
\label{figure:H76alpha_profiles}
\end{figure}

Quantitatively, we see that the predicted peak intensities under non-LTE conditions for apparently single-peaked RRL profiles agree very well with those reported in the literature (see Table \ref{table:peak_intensities}), with the exception of the only spatially-resolved cm-wavelength RRL to date, the H66$\alpha$. We predict its zero-intensity linewidth ($\Delta$v$\sim$100 km s$^{-1}$), although we clearly underestimate the peak intensity reported by Loinard \& Rodr\'iguez (2010) even for non-LTE emission (i.e. peak intensity of 9.0 mJy for the LTE case versus the 16.3 mJy predicted in the non-LTE case, and the 24.9 mJy observed by \cite{Loinard2010}). Thus, we conclude that even at this low frequency, the stimulated amplification of the radiation is still significant as was also claimed by Mart\'in-Pintado et al. (1993). This is also supported by the available RRL profile with the lowest frequency, the H76$\alpha$ line. Its asymmetric profile is a clear indication that stimulated emission plays a key role for this RRL (\cite{Escalante1989}). In this case, our non-LTE prediction matches quite well with the reported RRL profile, while the LTE prediction greatly underestimates its peak intensity and it is not able to explain the asymmetry in the profile (see Fig. \ref{figure:H76alpha_profiles}). The larger intensity of the blue-shifted peak is due to the existence of a larger background continuum emission to be amplified by the foreground ionized material approaching us. Thus, we conclude that even at 2 cm, the consideration of non-thermal emission is essential to explain the observations. 

\subsection{High-frequency radio-recombination lines. The inner disk: a distorted Keplerian-rotation}

For RRLs with quantum numbers $n\ge30$, our model accurately predicts both the peak intensities (see Table \ref{table:peak_separation}) and the maser peak-to-peak velocity separations, $(\mathrm{v}_\mathrm{red}-\mathrm{v}_\mathrm{blue})_\mathrm{obs}$, which does not depend so strongly on the physical conditions but on the kinematics of the region traced by the RRLs. Table \ref{table:peak_separation} shows that the predicted change of $(\mathrm{v}_\mathrm{red}-\mathrm{v}_\mathrm{blue})$ from H30$\alpha$ to H39$\alpha$ is consistent with the observations when one assumes the Storey \& Hummer departure coefficients. Since the maser peaks mainly trace the Keplerian-rotating ionized disk, we conclude that the derived central mass for the source is consistent with the reported $(\mathrm{v}_\mathrm{red}-\mathrm{v}_\mathrm{blue})_\mathrm{obs}$.

\begin{table*}
\caption{Comparison between the observed and predicted peak intensities and velocity peak-to-peak separations for Hn$\alpha$ RRLs toward MWC349A assuming a central mass of 38 $\mathrm{M}_\mathrm{\sun}$.}
\label{table:peak_separation}
\centering
\begin{tabular}{ccccccc} \hline
 Line& $\nu $&$S_{\mathrm{blue\ peak, obs}}$ &$S_{\mathrm{red\ peak, obs}}$&$S_{\mathrm{peak, pred}}$ &$(\mathrm{v}_\mathrm{red}-\mathrm{v}_\mathrm{blue})_\mathrm{obs}$ & $(\mathrm{v}_\mathrm{red}-\mathrm{v}_\mathrm{blue})_\mathrm{pred}$\\
 & $[\mathrm{GHz}]$& $ [\mathrm{Jy} ]$&$ [\mathrm{Jy}  ]$&$ [\mathrm{Jy}  ]$&$ [\mathrm{km\ s}^{-1} ]$ & $ [\mathrm{km\ s}^{-1}  ]$\\ \hline \hline
H39$\alpha$&106.74&0.75$^{\mathrm{a}}$&0.7$^{\mathrm{a}}$&0.7&$\sim$34$^{\mathrm{b}}$&30.5$\pm$0.1\\ 
H36$\alpha$&135.29&1.1$^{\mathrm{c}}$&1.5$^{\mathrm{c}}$&1.2&36.5$\pm$0.4$^{\mathrm{a}}$&35.8$\pm$0.5\\ 
H35$\alpha$&147.05&1.18$\pm$0.07$^{\mathrm{a}}$&2.48$\pm$0.07$^{\mathrm{a}}$&1.5&38.5$\pm$0.5$^{\mathrm{b}}$&36.4$\pm$0.5\\
H30$\alpha$&231.90&24.4$^{\mathrm{c}}$&38.9$^{\mathrm{c}}$&19.1&51.0$\pm$0.1$^{\mathrm{a}}$&51.0$\pm$0.1\\ 
H27$\alpha$&316.42&25.6$\pm$0.6$^{\mathrm{a}}$&34.4$\pm$0.4$^{\mathrm{a}}$&137&52.5$\pm$0.1$^{\mathrm{a}}$&54.4$\pm$0.1\\  
H26$\alpha$&353.62&39.7$\pm$0.4$^{\mathrm{a}}$&60.0$\pm$0.6$^{\mathrm{a}}$&873&51.8$\pm$0.1$^{\mathrm{a}}$&58.2$\pm$0.1\\  

H21$\alpha$&662.40&309$\pm$11$^{\mathrm{c}}$&261$\pm$11$^{\mathrm{c}}$&$\sim10^{14}$&50.1$\pm$0.4$^{\mathrm{c}}$&90.0$\pm$0.4\\ \hline 
\end{tabular}\
\begin{list}{}{}

\item[$^{\mathrm{a}}$] \cite{Thum1995}.
\item[$^{\mathrm{b}}$] \cite{Martinpintado1994}.
\item[$^{\mathrm{c}}$] \cite{Thum1994b}.
\end{list}
\end{table*}

The kinematics of the innermost region of the ionized circumstellar disk can only be studied by using the RRLs at higher frequencies since the continuum becomes optically thinner with increasing frequency, tracing the innermost parts. At present there are velocity-resolved observations for three transitions with frequencies higher than that of the H30$\alpha$. These are the H27$\alpha$, H26$\alpha$, and H21$\alpha$. Table \ref{table:peak_separation} shows the predicted peak intensity for the three maser lines are clearly overestimated in the unsaturated case, especially for the H21$\alpha$ which is roughly 10$^{14}$ times more intense than that observed (Thum et al. 1994b). Thus, as discussed, saturation effects could prevent the maser intensity from increasing with an exponential dependence.




Table \ref{table:peak_separation} shows that our modelling is not able to explain the observed maser peak-to-peak velocity separations, $(\mathrm{v}_\mathrm{red}-\mathrm{v}_\mathrm{blue})$. The predicted peak-to-peak velocity separation increases with the frequency of the RRL, as expected from the assumption that the disk follows a Keplerian rotation, while observations show that it remains approximately constant from the H30$\alpha$ to the H21$\alpha$ line.

This fact could be explained if the submm RRL maser emission occured in the same regions as for the H30$\alpha$. This would be the case if there was a lack of ionized material inside the region where the bulk of the H30$\alpha$ emission arises or an inner rarefication of the disk as claimed by \cite{Kraus2000}. Nevertheless, in such a case, it would not be possible to explain the radio-continuum spectral index of 0.62 up to the frequency of the H21$\alpha$ line. On the other hand, saturation effects of interaction  between masing transitions could explain that the maser lines are produced in the same region (\cite{Strelnitski1996b}). Hengel \& Kegel (2000) studied the effects of the radiation field on the electronic level population by using a radiative transfer model for a spherically symmetric source. They derived the ranges of electron density and quantum numbers where maser emission arises. While the optimum electron density values for unsaturated maser amplification are very sensitive to the quantum numbers of the RRLs (Fig. 6 in \cite{Strelnitski1996b}), the saturation of the maser lines counteract this behaviour such that the optimum electron density of the bulk of maser lines are contained in a small range of electron densities and quantum numbers (Fig. 4 in \cite{Hengel2000}). Thus, saturated maser lines would arise from regions with similar electron densities, i.e. from regions of similar radial distance and thus similar radial velocities. In such a case, the peak-to-peak velocity separation for saturated lines would not increase with the RRL frequency but would remain constant as observed for $n<30$. This provides good evidence that RRLs with $n<30$ are saturated and, therefore, our model is not able to reproduce these lines.

A third possibility to explain the constant peak-to-peak velocity separation from the H30$\alpha$ to submm RRLs results from the inner regions of the disk not following the Keplerian rotation of the outer parts. The kinematic discontinuity could likely be because of disturbances in the kinematics of the ionized circumstellar disk in the region where the outflow is launched.

\subsection{Fit of the Hn$\beta$ radio-recombination lines}

Finally, we have used the Hn$\beta$ transitions to further constrain the physical parameters derived from our modelling. The predicted Hn$\beta$ line intensities for LTE and non-LTE assumptions are shown in Table \ref{table:Hbeta_RRLs}. The physical structure and the departure coefficients assumed by MORELI predict a significant amplification for Hn$\beta$ lines, but much weaker than for the Hn$\alpha$ lines due to their lower $|\beta_{mn}|$. However, contrary to what was found in the Hn$\alpha$ transitions, the Hn$\beta$ transitions can be best explained by assuming LTE emission. The difference between the observed and predicted LTE intensities are within $\sim$4 times the rms noise of the data. The good fit obtained for the observed Hn$\beta$ transitions could represent another proof that supports the quantitative parameter values obtained by our modelling, but only if the Hn$\beta$ lines are not amplified.

\begin{figure}
\centering
\includegraphics[width=8cm]{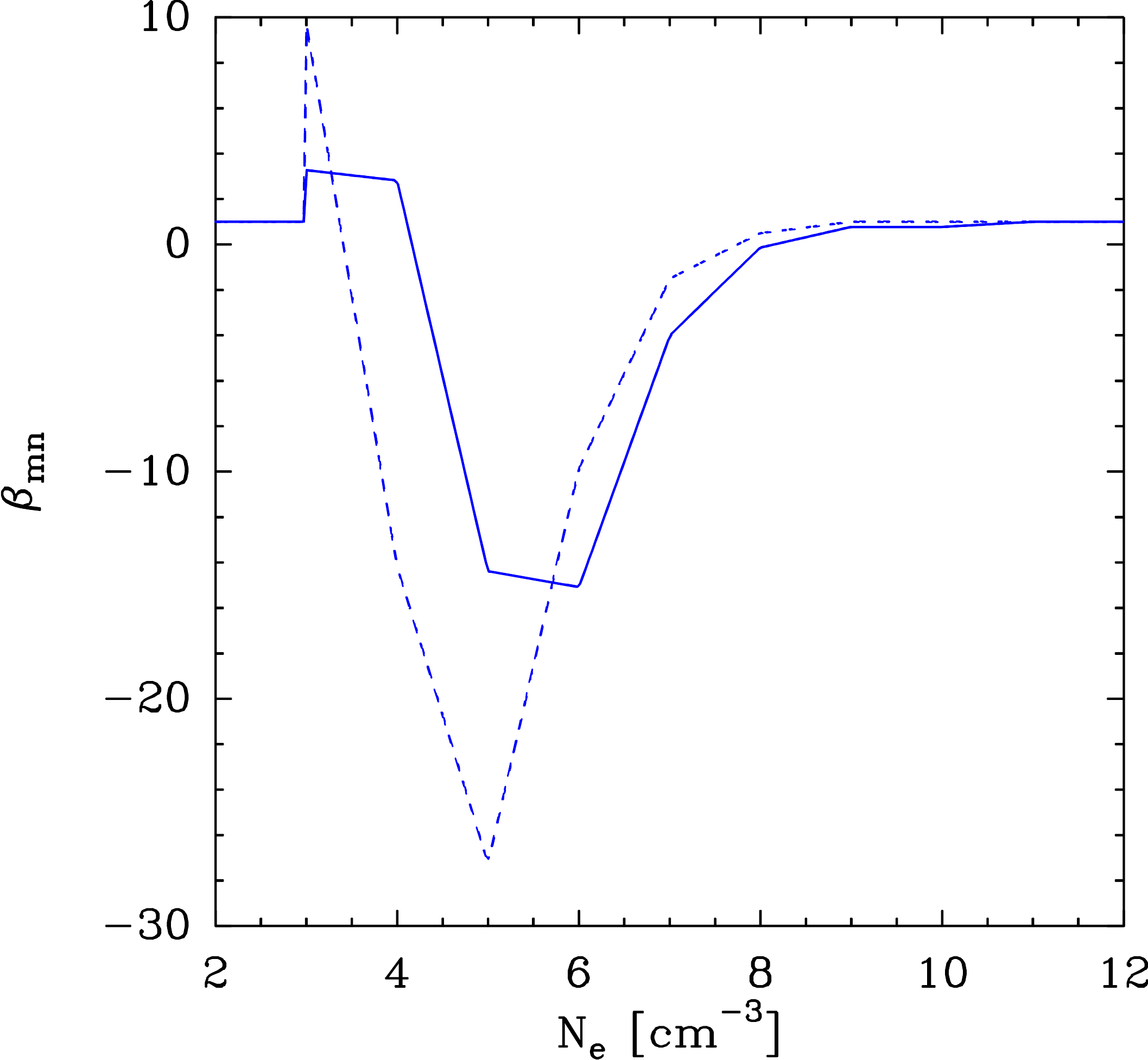}
\caption{Estimated $\beta_{mn}$ coefficients for an ionized gas with electron temperature of $10^4$ K for electron densities between $10^2$ and $10^{12}$ cm$^{-3}$. The solid and dashed lines show the $\beta_{mn}$ values derived by \cite{Storey1995} for H30$\alpha$ and H38$\beta$ transitions, respectively. We note that the H38$\beta$ lines thermalize at electron densities lower than for H30$\alpha$ lines.}
\label{beta_vs_Ne}
\end{figure}

However, the amplification predicted by our model for the Hn$\beta$ lines is unclear. A possible explanation could be related to the fact that RRLs sample the physical conditions where the continuum emission is optically thin. Thus, pairs of Hn$\alpha$ and Hn$^{\prime}\beta$ transitions with similar frequencies arise from regions with similar electron densities. For a certain electron density, the $b_{n}$ coefficients might represent roughly the electronic population for a range of quantum numbers (Hn$\alpha$ lines), but not for higher quantum numbers (Hn$^{\prime}\beta$ lines) for which the electronic levels are closer to thermalization. As an example, we can consider the H30$\alpha$ and H38$\beta$ pair at similar frequencies mainly sample regions with electron densities of $\sim10^7$ cm$^{-3}$. However, we can see in Fig. \ref{beta_vs_Ne} that the $\beta_{mn}$ coefficients for H30$\alpha$ and H38$\beta$ significantly differ. In particular, the $\beta_{mn}$ coefficients for H38$\beta$ are close to zero while for H30$\alpha$ we find $\beta_{mn}=-4$. As we can also see in Fig. \ref{beta_coefficients}, the numerical error in the estimation of the $\beta_{mn}$ coefficients increases when $\beta_{mn}$ gets close to thermalization. If the electronic levels actually thermalize faster when there is high electron density than when the model predictions are considered, then the $\beta_{mn}$  coefficients for the Hn$\beta$ lines would be closer to zero than our model assumes and the amplification would be negligible. In this case the H38$\beta$ line would appear to be emitted under LTE conditions as observed. Thus, the uncertainty for the $\beta_{mn}$ when the electronic population approaches thermalization might explain why the Hn$\beta$ are not amplified as expected from our model.



\begin{table*}
\caption{Comparison between the observed and predicted peak intensities for Hn$\beta$ RRLs toward MWC349A.}
\label{table:Hbeta_RRLs}
\centering
\begin{tabular}{cccccc}\hline
Line& $\nu$ &$S_\mathrm{peak, obs}$$^{\mathrm{a}}$ & $S_\mathrm{peak, LTE\ pred}$ &$\Delta S_{\mathrm{peak}}$$^{\mathrm{b}}$&$S_\mathrm{peak, non-LTE\ pred}$ \\
&$\left [ \mathrm{GHz} \right ]$& $\left [ \mathrm{mJy} \right ]$&$\left [ \mathrm{mJy} \right ]$&$\left [ \mathrm{mJy} \right ]$&$\left [ \mathrm{mJy} \right ]$\\ \hline \hline
H48$\beta$&111.89&68$\pm$10&52&1.5$\sigma$&96.7\\
H38$\beta$&222.01&171$\pm$10&211&4.3$\sigma$&433 \\
H33$\beta$&335.21&330$\pm$50&456&2.6$\sigma$&981 \\ 
H32$\beta$&366.65& $\sim$460$\pm$40&535&2.0$\sigma$&1186 \\ \hline 
\end{tabular}
\begin{list}{}{}
\item[$^{\mathrm{a}}$] \cite{Thum1995}.
\item[$^{\mathrm{b}}$] $\Delta S_{\mathrm{peak, LTE\ pred}}\equiv S_\mathrm{peak, obs}-S_\mathrm{peak, pred}$. Their values are shown in units of the rms noise of the spectra, $\sigma$. 
\end{list}
\end{table*}

\section{Conclusions}
\label{section_conclusions}



%








This paper describes the MORELI 3D radiative transfer model to predict RRL emission under LTE and non-LTE conditions in a variety of geometries, physical structures, and kinematics. Previous versions of this model were used in order to study different UC-HII regions: CRL618 (\cite{Martinpintado1988}), MWC349A (\cite{Martinpintado1989a}, \cite{Martinpintado2011}), Cepheus A HW2 (\cite{Jimenez-Serra2011}), and MonR2-IRS2 (\cite{Jimenez-Serra2013}).  In this paper, we used MORELI to explain the continuum and RRL data observed in MWC349A in a coherent way. This is the first complete modelling of MWC349A, which not only constrains the model parameters from a subset of RRL data but also for the whole set of observed RRL. Thus, our new results for the geometry, struture, and kinematics of MWC349A provide the best description of the characteristics of the ionized wind and outflow in this source, and of the central mass of the source.


Mainly on the basis of the explanation of the complex behaviour of the H30$\alpha$ centroid map and its RRL profile, we have strongly constrained the geometry, disk inclination, physical conditions and kinematics of MWC349A. These results were also derived with a systematic study of the whole set of radio-continuum and RRL observations. We found that the set of departure coefficients $b_n$ provided by \cite{Storey1995} describes much better the intensities of mm and submm masers than those given by \cite{Walmsley1990} for the physical structure and geometry derived for MWC349A. With these coefficients we are able to explain for the first time intensities of the maser spikes as well as the peak-to-peak separation between the spikes of mm RRLs. 


Thus, our modelling is supported by the fact that they explain the bulk of the different observational data available (SED, radio-continuum emission maps, RRL profiles and velocity-integrated line intensities for RRLs) at very different wavelengths (from mm to cm wavelengths). In addition, our model provides a general view of the dominant processes involved in the amplification of RRLs for the different frequency ranges. While maser amplification seems to dominate for Hn$\alpha$ RRLs with n$<41$, stimulated amplification is still important even up to frequencies as low as that of the H76$\alpha$, 15 GHz. However, our model does not reproduce the peak intensities, peak-to-peak velocity separation, and velocity-integrated line intensities for submm and infrared RRLs with n$<30$. Thus, one of the open issues to be addressed in the future is taking into account the saturation effects. High spatial resolution and velocity-resolved studies by using high-frequency RRLs (with $n<30$) are also required to establish the kinematics of inner regions.

The main difficulty found for our non-LTE modelling is to predict the observed peak intensities for the Hn$\beta$ line. While we expected the Hn$\beta$ lines to be out of LTE and with significant amplification, their observational peak intensities agree very well with the predictions for LTE emission. High spatial resolution observations of Hn$\beta$ should be perfomed in order to explain this puzzling behaviour. However, precise calculations of the $b_n$ coefficients for the geometry and physical conditions in MWC349A might explain these discrepancies.

In summary, we have used MORELI to better constrain the characteristics of both ionized components on MWC349A, disk and outflow, as proved by the large amount of observational data explained by our modelling. Further progress depends on observations which combine high angular and high velocity resolution, like those possible with ALMA.

                                                           ́

\begin{appendix}
\section{Gaunt factors}
\label{gaunt_appendix}
In this appendix we show the equations used by MORELI to estimate the Gaunt factors. In the literature we find different approximations for the free-free Gaunt factor for different frequency and temperature regimes. MORELI uses the analytic approximation given by \cite{Gronenschild1978}, which reproduces the computational results of \cite{Karzas1961} with an accuracy of 10 \% in the frequency and electron temperature ranges where $10^{-2}<h \nu / \left(k T_\mathrm{e}\right)<10^3$:

\begin{eqnarray}
&&g_{\mathrm{ff},\nu}=\sqrt{\left(\frac{\sqrt{3}}{\pi} \mathrm{e}^x K_0(x)\right)^2+\left(a-b\cdot \log{\left(\frac{h \nu}{k T_\mathrm{e}}\right)}\right)^2}\\
&&\gamma^2=Z^2\frac{h \nu_0}{k T_\mathrm{e}}\nonumber\\ 
&&x=\frac{1}{2}\frac{h \nu}{k T_\mathrm{e}}\left(1+\sqrt{10 \gamma^2}\right) \nonumber\\
&&a=1.20 \exp{\left[-\left(\frac{\log{\gamma^2}-1}{3.7}\right)^2\right]} \nonumber\\
&&b=0.37 \exp{\left[-\left(\frac{\log{\gamma^2}+1}{2}\right)^2\right]} \mathrm{,} \nonumber
\label{equation:Gaunt_factor}
\end{eqnarray}

\noindent where $K_o$ is the modified Bessel function of the second kind, $Z$ the effective charge of the ionized gas, and $\nu_0$ the hydrogen ionization frequency. This approximation is equivalent to that of \cite{Leitherer1991} in the Rayleigh-Jeans regime ($h\nu \ll k T_\mathrm{e}$) provided that the frequency exceeds the plasma resonance frequency, $\nu\gg\nu_\mathrm{p}$ (where $\nu_\mathrm{p} = N_\mathrm{e} e^2 / \left(\pi m_\mathrm{e}\right)^{1/2}$). Thus, the Gronenschild \& Mewe expression can be used in a broader range of frequencies than the previous one.







\begin{figure}
\centering
\includegraphics[width=8cm]{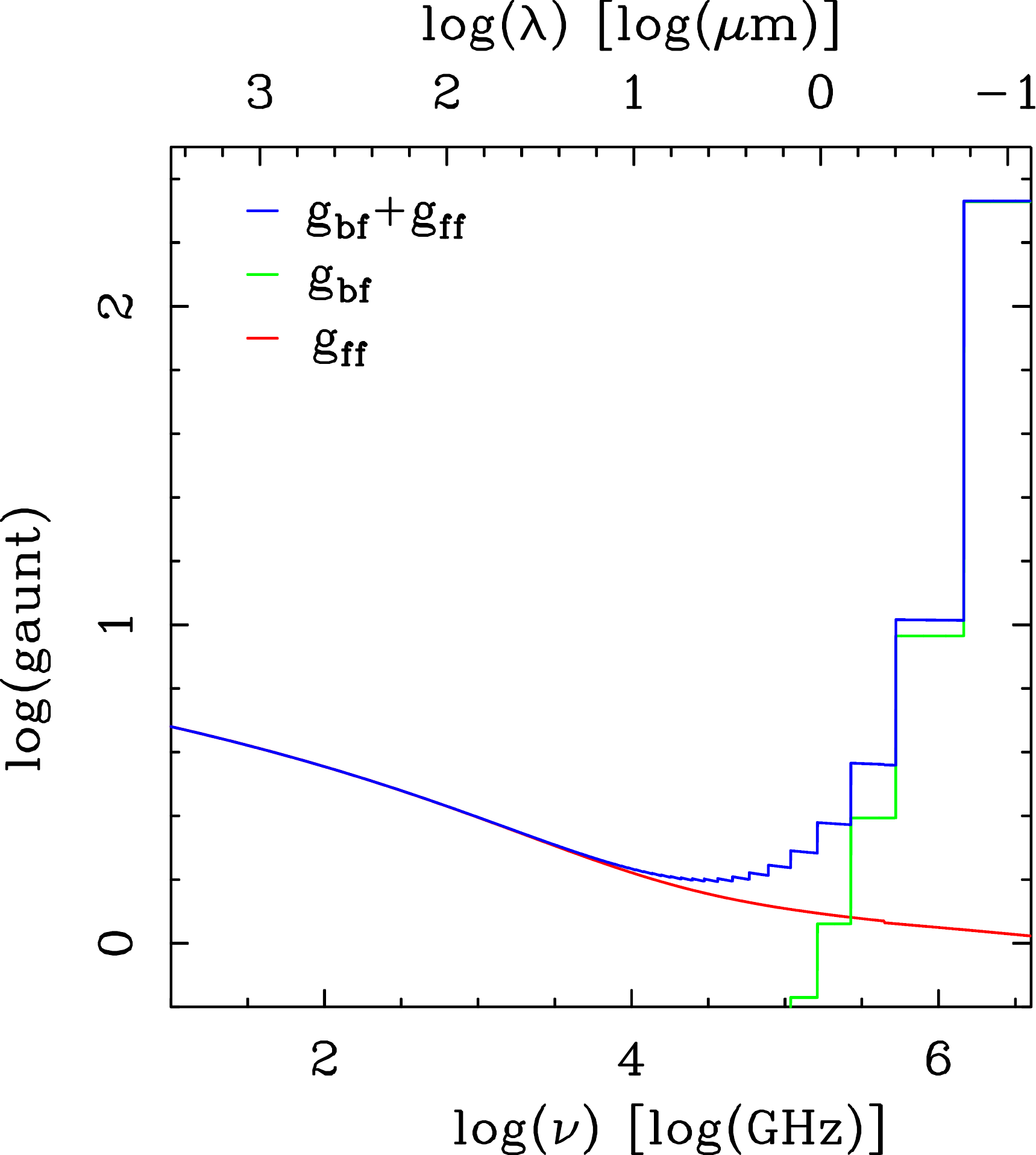}
\caption{Gaunt factor values for free-free and bound-free transitions and total Gaunt factor ($g_{\mathrm{ff}}$, $g_{\mathrm{fb}}$ and $g_{\mathrm{ff}}+g_{\mathrm{fb}}$ respectively) for frequencies between 10 and $4\times10^6$ GHz. We assumed a temperature of 10000 K.}
\label{figure:gaunt}
\end{figure}

In the absence of dust, free-free continuum emission is the dominant process for wavelengths $\lambda \gtrsim 10\ \mathrm{\mu m}$, while at NIR and optical wavelengths for which $g_{\mathrm{bf},\nu}\simeq1$, the bound-free processes become significant. This is shown in Fig. \ref{figure:gaunt}, where we plot the relative contribution of the bound-free and free-free Gaunt factors. The bound-free Gaunt factor is calculated by using the expression given by \cite{Brussaard1962}


\begin{eqnarray}
&&g_{\mathrm{bf},\nu} = 2\Theta \sum_{n=m}^{\infty} g_n(\nu)\frac{\mathrm{e}^{\Theta/n^2}}{n^3}\\
&&\Theta = \frac{h \nu_0}{k T_e} \nonumber \\
&&m= int\left(\sqrt{\frac{\nu_0}{\nu}}\right)+1 \mathrm{,} \nonumber
\end{eqnarray}

\noindent where $g_n(\nu)$ is approximated to 1 for all the frequencies. This approximation yields results with an accuracy of $\sim$ 10-20\%. 

\end{appendix}

\begin{appendix}
\section{Radio-recombination line profiles}
\label{RRL_profiles}


The RRL profiles of UC-HII regions are due to the combination of different broadening mechanisms whose convolution gives the observed profile. In Sect. \ref{kinematics_modelling} we have shown the large-scale motions that give as a result the broadening of the profile. However, even in every cell of the model (Sect. \ref{section_model}), there is a profile broadening caused by the microscopic motions of the ionized gas, such as turbulent and thermal motions.

Regarding the thermal motions, one can consider the approximation that the ionized gas is well described by the Maxwell velocity distribution. It subsequently produces a Gaussian profile because of the Doppler shifting. Likewise, the turbulent motions can be also described by a Gaussian distribution governed by a parameter known as turbulent velocity, $\mathrm{v}_\mathrm{tu}$. Thus, the velocity half-width at half-maximum height, $\Delta \mathrm{v}_\mathrm{g}$, of the Gaussian profile due to both contributions can be estimated using the expression (\cite{Mezger1967})

\begin{equation}
\Delta \mathrm{v}_\mathrm{g}=2\sqrt{\ln{(2)}}\sqrt{\frac{2k T_\mathrm{e}}{m_\mathrm{e}}+\frac{2}{3}v_\mathrm{tu}^2} \mathrm{.}
\end{equation}

Additionally, there is a contribution due to the broadening of the electronic levels in regions with high electron density. This is known as pressure broadening and it produces a Lorentzian profile. This broadening can be specified by the half-width at half-maximum height, $\Delta \mathrm{v}_\mathrm{l}$, which has been computed numerically using analytic expressions that approximate the results derived from classical and semi-classical theory (\cite{Gee1976}) for the different quantum number ranges of the RRLs (see Table \ref{table:pressure_broadening}).


\begin{table}
\caption{Analytic approximations of the pressure broadening for different ranges of electronic levels.}
\label{table:pressure_broadening}
\begin{tabular}{cc}\hline
Range of electronic levels& Pressure broadening, $\Delta\nu_\mathrm{l}$\\ \hline

$n>100$&$1.9\times10^{-8} n^{4.4} N_\mathrm{e}/\left(Z^2 T_\mathrm{e}^{0.1}\right)\ \ ^{\mathrm{a}}$\\
$30<n<100$&$6.7\times10^{-9} n^{4.6} N_\mathrm{e}/\left(Z^2 T_\mathrm{e}^{0.1}\right)\ \ ^{\mathrm{b}}$\\
$n<30$&$8.0\times10^{-10} n^{5} N_\mathrm{e}/\left(Z^2 T_\mathrm{e}^{0.1}\right)\ \ \ ^{\mathrm{c}}$\\ \hline
\end{tabular}
\begin{list}{}{}
\item[$^{\mathrm{a}}$] \cite{Brocklehurst1972}.
\item[$^{\mathrm{b}}$] \cite{Walmsley1990}.
\item[$^{\mathrm{c}}$] Strelnitski et al. (1996b).
\end{list}
\end{table}

The convolution of the aforementioned Gaussian and Lorentzian distributions results in a Voigt profile, which can be expressed as a function of $\Delta\nu_\mathrm{l}$ and $\Delta\nu_\mathrm{g}$ (where $\Delta\nu_\mathrm{g}=\frac{\nu}{c}\Delta \mathrm{v}_\mathrm{g}$) by


\begin{equation}
\Phi_{\nu} \left(\Delta\nu_\mathrm{l},\Delta\nu_\mathrm{g}\right)=\frac{1}{\sqrt{\pi}\Delta\nu_\mathrm{g}}\int_{-\infty}^{\infty}\frac{\Delta\nu_\mathrm{l}/\pi}{\Delta\nu_\mathrm{l}^2+\left(\nu-\nu'\right)^2} \mathrm{e}^{-\left(\nu' / \Delta\nu_\mathrm{g}\right)^2}d\nu' \mathrm{.}
\end{equation}

This integral is numerically computed in MORELI following the formulation of \cite{Kielkopf1973} to describe the RRL emission in every cell of the source. Besides the intrinsic line profile, the projection of the large-scale motions (modelled in Sect. \ref{kinematics_modelling}) along the line-of-sight gives rise to a dynamical broadening of the different RRL profiles. 

\end{appendix}

\begin{appendix}
\section{Effective radius for an optically thin wind}
\label{appendix_effective_radius}
The total free-free continuum flux for a spherical outflow in the circular region in the plane of the sky of radius $R_{\mathrm{eff}}$ is given by


\begin{eqnarray}
\bf{F_{\nu}}&=& 2 \pi r_{\mathrm{min}}^2 \int_{0}^{R_{\mathrm{eff}}} B_{\nu}(T_{\mathrm{e}}) \left(1-\mathrm{e}^{-\tau_\nu(\rho)}\right) \rho \, \mathrm{d} \rho\\
&=& 2 \pi r_{\mathrm{min}}^2 \Bigg{[} \int_{0}^{1} B_{\nu}(T_{\mathrm{e}}) \left(1-\mathrm{e}^{-\tau_\nu(\rho)}\right) \rho \, \mathrm{d} \rho \nonumber \\
&\ &\ \ \ \ \ + \int_{1}^{R_{\mathrm{eff}}} B_{\nu}(T_{\mathrm{e}}) \left(1-\mathrm{e}^{-\tau_\nu(\rho)}\right) \rho \, \mathrm{d} \rho\Big{]}\ \mathrm{,} \nonumber
\label{equation:intensidad}
\end{eqnarray}


\noindent where $\rho$ and $R_{\mathrm{eff}}$ are in units of the radius of the central hole without ionized gas, $r_\mathrm{min}$, with $\rho=\sqrt{x^2+y^2}$ and $R_{\mathrm{eff}}>1$.

The optical depth along the line-of-sight located at a radius $\rho$ from the center of the source, $\tau_\nu\left(\rho\right)$, for the spherical wind isotropically expanding at constant velocity, is given by

\begin{eqnarray}
\tau_\nu \left({\rho}\right)&=& \frac{\pi}{2 \rho^3} r_{\mathrm{min}} \ \kappa_{\nu}(\mathrm{r}=1)\ \ \  \mathrm{if}\ \rho>r_\mathrm{min} \\
\tau_\nu \left({\rho}\right)&=& \frac{1}{\rho^3} r_{\mathrm{min}} \Big{[} \frac{\pi}{2} - \rho\sqrt{1-\rho^2}\nonumber  \\ 
&-&\arctan{\left(\frac{\sqrt{1-\rho^2}}{\rho} \right)} \Big{]} \kappa_{\nu}(\mathrm{r}=1)\ \ \    
\mathrm{if}\ \rho<r_\mathrm{min} \nonumber
\end{eqnarray}

By using the optical depths given above, the integration of Eq. \ref{equation:intensidad} in a circumference of radius $R_{\mathrm{eff}}$ for the optically thin case, $1-\mathrm{e}^{-\tau_\nu(\rho)}\approx \tau_\nu\left(\rho\right)$, results as

\begin{eqnarray}
\label{intensidad_opt_thin}
\bf{F_{\nu}}&=& \pi r_{min}^3 B\left(\nu\right) \kappa_{\nu}(\mathrm{r}=1) \Big{[} \pi \left(1-\frac{1}{R_{\mathrm{eff}}}\right)\\ \nonumber
&+& \int_{0}^{1} \frac{1}{\rho^2} \left(\pi - 2\rho\sqrt{1-\rho^2} - 2\arctan{\left(\frac{\sqrt{1-\rho^2}}{\rho}\right)}\right)  \, \mathrm{d} \rho\Big{]}\nonumber \\
&=&\pi r_{min}^3 B\left(\nu\right) \kappa_{\nu}(\mathrm{r}=1) \left[ \pi \left(1-1/R_{\mathrm{eff}}\right) + 4-\pi\right] \mathrm{.} \nonumber
\end{eqnarray}
If we define $R_{\mathrm{eff}}$ as the radius containing a given percentage, $p$, of the total continuum flux, it results from Eq. \ref{intensidad_opt_thin} the following effective radius in units of $r_{\mathrm{min}}$:

\begin{equation}
\label{radio_efectivo}
R_{\mathrm{eff}}=\frac{\pi}{4\left(1-p\right)}
\end{equation}
\end{appendix}

\begin{appendix}

\section{Kinematics}
\label{kinematics_section}
\subsection{Disk kinematics}
\label{disk_kinematics}

In this appendix, we derive Eq. \ref{vz_equation} (Sect. \ref{outflow_kinematics_subsection}) describing the velocity component along the $z$ axis for a Keplerian rotating ionized disk. We assume that the only disk velocity component is the rotation around the revolution axis specified by the $x_{\mathrm{d}}$ axis. Thus, in order to describe its velocity, the cylindrical polar system of coordinates unit vectors $(\vec{e_{\rho}},\vec{e_{\varphi}},\vec{e_{x_{\mathrm{d}}}})$ and coordinates $(\rho,\varphi,x_{\mathrm{d}})$ is the most suitable. In this case the velocity is given by 

\begin{equation}
\vec{v}=r \dot{\varphi}\vec{e_{\varphi}} \mathrm{.}
\label{equation_v}
\end{equation}

Describing Eq. \ref{equation_v} in a Cartesian coordinate system with the same polar axis and unit vector $(\vec{y_{d}},\vec{z_{\mathrm{d}}},\vec{x_{\mathrm{d}}})$ \footnote{The $\vec{e_{\phi}}$ unit vector is described in Cartesian coordinates as follows: $\vec{e_{\phi}}=-\sin(\varphi) \vec{e_{x_d}}+\cos(\varphi) \vec{e_{y_d}}$.}, and assuming a Keplerian rotation velocity, Eq. \ref{equation_v} results as 

\begin{equation}
\vec{v}=r \dot{\varphi}\vec{e_{\varphi}}=\frac{V_\mathrm{Kepler}}{(y_{\mathrm{d}}+z_{\mathrm{d}})^{1/4}}(-\sin(\varphi)\vec{e_{y_\mathrm{d}}}+\cos(\varphi)\vec{e_{z_\mathrm{d}})} \mathrm{,}
\label{v_disk_cartesian_coord}
\end{equation}

\noindent where $V_\mathrm{Kepler}=\sqrt{G M}$.

Since in the general case, the disk could be tilted with respect to the line-of-sight with an angle $\theta_\mathrm{i}$ (see Fig. \ref{figure:figura}), we apply a rotation of coordinates around the axis $y_{\mathrm{d}}$ to lead to the Cartesian coordinate system with unit vectors $(\vec{e_\mathrm{y}},\vec{e_\mathrm{z}},\vec{e_\mathrm{x}})$ orientated as described in Sect. \ref{section_model}. Such a coordinate transformation is given by the equations

\begin{eqnarray}
&&\vec{e_{z_{\mathrm{d}}}}=\cos(\theta_i)\vec{e_z}-\sin(\theta_i)\vec{e_x} \mathrm{,} \nonumber\\ 
&&\vec{e_{x_{\mathrm{d}}}}=\sin(\theta_i)\vec{e_z}+\cos(\theta_i)\vec{e_x} \mathrm{,} \nonumber \\
&&\vec{e_{y_{\mathrm{d}}}}=\vec{e_y} \mathrm{.}
\end{eqnarray}

In this coordinate system, Eq. \ref{v_disk_cartesian_coord} is

\begin{eqnarray}
&&\vec{v}=v\left [-\sin(\varphi)\vec{e_{y}}+\cos(\varphi)\cos(\theta_i)\vec{e_z}-\cos(\varphi)\sin(\theta_\mathrm{i})\vec{e_x} \right ] \\
&&v=\frac{V_\mathrm{Kepler}}{(y_{\mathrm{d}}^2+z_{\mathrm{d}}^2)^{\frac{1}{4}}} \mathrm{,} \nonumber 
\label{v_cartesian_coord}
\end{eqnarray}

\noindent where $\cos(\varphi)=\frac{y_{\mathrm{d}}}{\sqrt{y_{\mathrm{d}}^2+{z_\mathrm{d}}^2}}\ $, $y_\mathrm{d}=y\ $, and $z_{\mathrm{d}}=\cos(\theta_i) z-\sin(\theta_i)x$.

Thus, the velocity component along the line-of-sight (the $z$ axis) yields

\begin{equation}
v_z=V_\mathrm{Kepler}\frac{y \cdot \cos(\theta_i)}{(y^2+z_{\mathrm{d}}^2)^{3/4}} \mathrm{.}
\label{vz_sin_aproximacion}
\end{equation}

For a nearly edge-on disk ($\theta_i\ll 1$), Eq. \ref{vz_sin_aproximacion} can be approximated as

\begin{equation}
v_z=V_\mathrm{Kepler}\frac{y \cdot \cos(\theta_i)}{(y^2+z^2)^{3/4}} \mathrm{.}
\end{equation}




\subsection{Outflow acceleration}
\label{outflow_accel}

In this appendix, we derive Eq. \ref{vz_equation_with_accel} (Sect. \ref{outflow_kinematics_subsection}) describing the velocity component along the $z$ axis for an outflow expanding radially away from the central star reaching a terminal velocity of $v_0$ at a radius $r_{\mathrm{a}}$. The total radial velocity of the ionized gas at $r<r_{\mathrm{a}}$ is given by $v_{\mathrm{r}}=\frac{r}{r_{\mathrm{a}}}v_0$. Thus, its projection along the $z$ axis of the Cartesian coordinate system with the unit vector $(\vec{y_{d}},\vec{z_{\mathrm{d}}},\vec{x_{\mathrm{d}}})$ described in \ref{disk_kinematics} is

\begin{eqnarray}
\vec{v}&=&v_r \left (\sin(\theta)\cos(\varphi) \vec{e_{y_\mathrm{d}}}+\sin(\theta)\sin(\varphi) \vec{e_{z_\mathrm{d}}}+\cos(\theta)\vec{e_{x_\mathrm{d}}} \right ) \\
&=&v_0 \left (\frac{y}{r_\mathrm{a}} \vec{e_{y_\mathrm{d}}}+\frac{z}{r_\mathrm{a}} \vec{e_{z_\mathrm{d}}}+\frac{x}{r_\mathrm{a}}\vec{e_{x_\mathrm{d}}} \right ) \mathrm{,} \nonumber
\end{eqnarray}

\noindent where $\theta$ is the polar angle and $\varphi$ the azimuthal angle. 

Thus, we obtain the velocity projected to the $z$-axis 

\begin{equation}
v_z=v_0 \frac{z}{r_{\mathrm{a}}} \mathrm{.}
\end{equation}



\end{appendix}

\begin{acknowledgements} 
This work has been partially funded by MICINN grants AYA2010-21697-C05-01, FIS2012-39162-C06-01, and Astro-Madrid (CAM S2009/ESP-1496). Alejandro B\'aez-Rubio acknowledges support from grant JAE predoct (2009), CSIC, Spain. We would also like to thank Josefina Torres for providing Fig. \ref{figure:figura} and  L.F. Rodr\'iguez for providing the observational radio-continuum images of MWC349A (Fig. \ref{figure:radio_continuum}) and the H76$\alpha$ RRL profile (Fig. \ref{figure:H76alpha_profiles}). Finally, we are also grateful to the anonymous referee and Malcom Wamsley for very valuable comments.
\end{acknowledgements}


\begin{thebibliography}{}

\bibitem[Allen (1973)]{Allen1973} Allen, D.A.\ 1973, MNRAS, 161, 145 

\bibitem[Andrillat et al. 1996]{Andrillat1996} Andrillat, Y., Jaschek, M. \& Jaschek, C. 1996, A\&AS, 118, 495 

\bibitem[Aitken et al. 1990]{Aitken1990} Aitken, D.K., Smith, C.H., Roche, P.F. et al. 1990, MNRAS, 247, 466 

\bibitem[Altenhoff et al. (1994)]{Altenhoff1994} Altenhoff, 
W.\ J., Thum, C. \& Wendker, H.\ J.\ 1994, A\&A, 281, 161 

\bibitem[Beichman (1988)]{Beichman1988} Beichman, C.A., Neugebauer, G., Habing, H.J. et al. \ 1988, Infrared Astronomical Satellite (IRAS) Catalogs and Atlases, Vol. 1: Explanatory Supplement, NASA Washington, DC)

\bibitem[Blandford \& Payne 1982]{Blandford1982} Blandford, R. D. \& Payne, D. G.\ 1982, MNRAS, 199, 883  

\bibitem[Braes et al. 1972]{Braes1972} Braes, L.L.E., Habing, H.J. \& Schoenmaker, A.A. 1972, Nature, 240, 230

\bibitem[Brocklehurst \& Seaton (1972)]{Brocklehurst1972} Brocklehurst, M. \& Seaton, M.J. 1972, MNRAS, 157, 179

\bibitem[Brussaard  \& Van de Hulst (1962)]{Brussaard1962} Brussaard, P.J. \& Van de Hulst, H.C. 1962, Rev. Mod. Phys., 34, 507


\bibitem[Cohen et al. 1985]{Cohen1985} Cohen, M., Bieging, J. H., Welch, W. J. et al. 1985, ApJ, 292, 249

\bibitem[Cox et al. 1995]{Cox1995} Cox, P., Mart\'in-Pintado, J., Bachiller, R. et al.  1995, A\&A, 295, L39

\bibitem[Churchwell 1990]{Churchwell1990} Churchwell, E. 1990, A\&A Rev, 2, 79 

\bibitem[Danchi et al. 2001]{Danchi2001} Danchi, W.C., Tuthill, P.G. \& Monnier, J.D. 2001, ApJ, 562, 440 

\bibitem[Davies et al. 2010]{Davies2010} Davies, B., Lumsden, S.L., Hoare, M.G. et al. 2010, MNRAS, 402, 1504 

\bibitem[Dupree \& Goldberg 1970]{Dupree1970} Dupree, A.K. \& Goldberg, L. 1970, ARA\&A, 8, 231

\bibitem[Einstein 1916]{Einstein1916} Einstein, A. 1916, Verh. Deutsch Phys. Ges. 18, 318

\bibitem[Escalante et al. 1989]{Escalante1989} Escalante, V., Rodr\'iguez, L. F., Moran, J. M. et al. 1989, RevMexAA, 17, 11 


\bibitem[Gee et al. 1976]{Gee1976} Gee, C.S., Percival, I.C., Lodge, J.G., \& Richards, D. 1976, MNRAS, 175, 209

\bibitem[Goldberg 1966]{Goldberg1966} Goldberg, L.\ 1966, ApJ, 144, 1225  

\bibitem[Gronenschild \& Mewe (1978)]{Gronenschild1978} Gronenschild, E.H.B.M. \& Mewe, R. 1978, A\&AS, 32, 283  



\bibitem[Hamann \& Simon 1988]{Hamann1988} Hamann, F. \& Simon, M. 1988, ApJ, 327, 876  

\bibitem[Harvey et al. (1979)]{Harvey1979} Harvey, P.M., Thronson, H.A. \& Gatley, I. 1979, ApJ, 231, 115

\bibitem[Hengel \& Kegel 2000]{Hengel2000} Hengel, C. \& Kegel, W.H. 2000, A\&A, 361, 1169

\bibitem[Hollenbach et al. 1994]{Hollenbach1994} Hollenbach, D., Johnstone, D., Lizano, S. et al. 1994, ApJ, 428, 654  


\bibitem[Jaffe \& Mart{\'i}n-Pintado 1999]{Jaffe1999} Jaffe, D. T. \& Mart{\'i}n-Pintado, J.\ 1999, ApJ, 520, 162 

\bibitem[Jim\'enez-Serra et al. 2007]{Jimenez-serra2007} Jim\'enez-Serra, I., Mart\'in-Pintado, J., Rodr\'iguez-Franco, A. et al. 2007, ApJ, 661, L187 

\bibitem[Jim\'enez-Serra et al. 2009]{Jimenez-Serra2009} Jim\'enez-Serra, I., Mart\'in-Pintado, J., Caselli, P. et al. 2009, ApJ, 703, L157

\bibitem[Jim\'enez-Serra et al. 2011]{Jimenez-Serra2011} Jim\'enez-Serra, I., Mart\'in-Pintado, J., B\'aez-Rubio, A. et al. 2011, ApJ, 732, L27

\bibitem[Jim\'enez-Serra et al. 2013]{Jimenez-Serra2013} Jim\'enez-Serra, I., B\'aez-Rubio, A., Rivilla, V.M. et al. 2013, ApJ, 764, L4

\bibitem[Karzas \& Latter (1961)]{Karzas1961} Karzas, W.J. \& Latter, R. 1961, ApJS, 6, 167


\bibitem[Kielkopf (1973)]{Kielkopf1973} Kielkopf, J.F. 1973, J. Opt. Soc. Amer. B., 63, 987

\bibitem[Kraus et al. 2000]{Kraus2000} Kraus, M., Kr\"ugel, E., Thum, C. \& Geballe, T.R. 2000, A\&A 362, 158

\bibitem[Kraus \& Lamers 2003]{Kraus2003} Kraus, M. \& Lamers, H.J.G.L.M. 2003, A\&A 405, 165

\bibitem[Kraus (2009)]{Kraus2009} Kraus, M. 2009, A\&A 494, 253 

\bibitem[Kraus et al. 2010]{Kraus2010} Kraus, S., Hofmann, K., Menten, K.M. et al. 2010, Nature 466, 339



\bibitem[Lamers et al. 1998]{Lamers1998} Lamers, H.J.G.L.M., Zickgraf, F.-J., de Winter, D. et al. 1998, A\&A 340, 117


\bibitem[Lee (1970)]{Lee1970} Lee, T.A.\ 1970, PASP, 82, 765 

\bibitem[Leitherer \& Robert (1991)]{Leitherer1991} Leitherer, C., Robert, C. 1991, A\&A 377, 629

\bibitem[Loinard \& Rodr\'iguez 2010]{Loinard2010} Loinard, L., \& Rodr\'iguez, L. F. 2010, ApJ, 722, L100

\bibitem[Lugo et al. (2004)]{Lugo2004} Lugo, J., Lizano, S. \& Garay, G. 2004, ApJ, 614, 807


\bibitem[Marston \& McCollumn 2008]{Marston2008} Marston, A.P. \& McCollum, B. 2008, A\&A, 477, 193

\bibitem[Mart\'in-Pintado et al. 1988]{Martinpintado1988} Mart\'in-Pintado, J., Bujarrabal, V., Bachiller, R. et al. 1988, A\&A, 197, L15 

\bibitem[Mart\'in-Pintado et al. 1989a]{Martinpintado1989a} Mart\'in-Pintado, J., Bachiller, R.,  Thum, C. et al. 1989a, A\&A, 215, L13

\bibitem[Mart\'in-Pintado et al. 1989b]{Martinpintado1989b} Mart\'in-Pintado, J., Bachiller, R. \&  Thum, C. 1989b, A\&A, 222, L9

\bibitem[Mart\'in-Pintado et al. 1993]{martinpintado1993} Mart\'in-Pintado,  J., Gaume, R., Bachiller, R. et al. 1993, ApJ, 418, L79 

\bibitem[Mart\'in-Pintado et al. (1994)]{Martinpintado1994} Mart\'in-Pintado, J., Neri, R., Thum, C. et al. 1994, A\&A, 286, 890 



\bibitem[Mart\'in-Pintado et al. 2011]{Martinpintado2011} Mart\'in-Pintado, J., Thum, C., Planesas, P. et al. 2011, A\&A, 530, L15


\bibitem[Meyer et al. 2002]{Meyer2002} Meyer, J.M., Nordsieck, K.H. \& Hoffman, J.L. 2002, AJ, 123, 1639

\bibitem[Meynet et al. 1994]{Meynet1994} Meynet, G., Maeder, A., Schaller, G. et al. 1994, A\&AS, 103, 97

\bibitem[Mezger \& Hoglund 1967]{Mezger1967} Mezger, P.G. \& Hoglund, B. 1967, ApJ, 147, 490

\bibitem[Miroshnichenko et al. 2005]{Miroshnichenko2005} Miroshnichenko, A.S., Bjorkman, K. S, Grosso, M. et al. 2005, A\&A, 436, 653

\bibitem[Nielbock et al. 2007]{Nielbock2007} Nielbock, M., Chini, R. \& Hoffmeister, V.H. 2007, ApJ, 656, L81

\bibitem[Olnon 1975]{Olnon1975} Olnon, F. M. 1975, A\&A, 39, 217

\bibitem[Osterbrock 1989]{Osterbrock1989} Osterbrock, D. E. 1989, Astrophysics of Gaseous Nebulae and Active galactic Nuclei (Mill Valley: University Science Books)  

\bibitem[Panagia \& Felli 1975]{Panagia1975} Panagia, N. \& Felli, M. 1975, A\&A, 39, 1

\bibitem[Planesas et al. (1992)]{Planesas1992} Planesas, P., Mart\'in-Pintado, J. \& Serabyn, E.\ 1992, ApJ, 386, L23  

\bibitem[Preibisch et al. 2011]{Preibisch2011} Preibisch, T., Ratzka, T., Gehring, T. et al. 2011, A\&A, 530, A40

\bibitem[Rodr\'iguez \& Bastian 1994]{Rodriguez1994} Rodr\'iguez, L.F. \& Bastian, T.S. 1994, ApJ, 428, 324  


\bibitem[Sandell et al. (2011)]{Sandell2011} Sandell, G., Weintraub, D.A., Hamidouche, M. 2011, ApJ, 727, 26

\bibitem[Schaller et al. 1992]{Schaller1992} Schaller, G., Schaerer, D., Meynet, G. et al. 1992, A\&AS, 96, 269

\bibitem[Schwartz (1980)]{Schwartz1980} Schwartz, P.R. \ 1980, PASP, 92, 534

\bibitem[Shu et al. 1994]{Shu1994} Shu, F., Najita, J., Ostriker, E. et al. 1994, ApJ, 429, 781

\bibitem[Simon \&  Dyck (1977)]{SimonDyck1977} Simon, T \& Dyck, H.M.\ 1977, AJ, 82, 725 


\bibitem[Storey \& Hummer (1995)]{Storey1995} Storey, P.J. \& Hummer, D. G. 1995, MNRAS, 272, 41

\bibitem[Strelnitski et al. 1996a]{Strelnitski1996a} Strelnitski, V., Haas, M.R., Smith, H.A. et al. 1996a, Science, 272, 1459

\bibitem[Strelnitski et al. 1996b]{Strelnitski1996b} Strelnitski, V. S., Ponomarev, V. O., \& Smith, H. A. 1996b, ApJ,470, 1118

\bibitem[Tafoya et al. 2004]{Tafoya2004} Tafoya, D., G\'omez, Y. \& Rodr\'iguez, L. F.\ 2004, ApJ, 610, 827 

\bibitem[Thum et al. 1992]{Thum1992} Thum, C., Mart\'in-Pintado, J. \& Bachiller, R.\ 1992, A\&A, 256, 507 



\bibitem[Thum et al. 1994a]{Thum1994a}Thum, C., Matthews, H. E., Mart\'in-Pintado, J. et al. 1994a, A\&A, 283, 582 

\bibitem[Thum et al. (1994b)]{Thum1994b} Thum, C., Matthews, H. E., Harris, A. I. et al. 1994b, A\&A, 288, L25 

\bibitem[Thum et al. (1995)]{Thum1995} Thum, C., Strelnitski, V. S., Marti\'n-Pintado, J. et al. 1995, A\&A, 300, 843 

\bibitem[Thum et al. 1998]{Thum1998} Thum, C., Mart\'in-Pintado, J., Quirrenbach, A. et al. 1998, A\&A, 333, L63 


\bibitem[Towle et al. 1999]{Towle1996} Towle, J.P., Feldman, P.A. \& Watson, J.K.G. 1996, ApJS, 107, 747  

\bibitem[Walmsley (1990)]{Walmsley1990} Walmsley, C.M. 1990, A\&A, 82, 201


\bibitem[White \& Becker 1985]{White1985} White, R.L. \& Becker, R.H. 1985, ApJ, 297, 677


\bibitem[Yudin 1996]{Yudin1996} Yudin, R.V. 1996, A\&A, 312, 234

\bibitem[Zinnecker \& Yorke 2007]{Zinnecker2007} Zinnecker, H. \& Yorke, H.W. \ 2007, ARA\&A, 45, 481










\end{thebibliography}
\end{document}